\documentclass[fleqn,usenatbib]{mnras}
\usepackage{newtxtext,newtxmath}
\usepackage[T1]{fontenc}
\usepackage{ae,aecompl}
\usepackage{graphicx}	
\usepackage{amsmath}	
\usepackage{threeparttable}
\usepackage{pgf,tikz}
\usepackage{bm,color,ulem}
\usepackage{graphicx}
\usepackage{epstopdf}
\usepackage{xcolor}
\usepackage{float}

\definecolor{mygray}{gray}{0.5}
\definecolor{deepgreen}{RGB}{20, 130, 22}
\definecolor{peach}{RGB}{180,70,100}
\definecolor{red2}{RGB}{250,0,200}

\newcommand{\der}{{\rm d}}

\newcommand{\blr}{{\bm {\hat l}}_{\rm r}}
\newcommand{\bl}{{\bm {\hat l}}}
\newcommand{\ve}{{\bm e}}
\newcommand{\Ir}{I_{\rm r}}
\newcommand{\Omr}{\Omega_{\rm r}}

\newcommand{\om}{\omega}
\newcommand{\Om}{\Omega}
\newcommand{\dg}{\delta}
\newcommand{\Dg}{\Delta}

\newcommand{\Msun}{{\rm M}_\odot}
\newcommand{\Rsun}{{\rm R}_\odot}

\newcommand{\be}{\begin{equation}}
\newcommand{\ee}{\end{equation}}

\title[Constraining the tilt in KH 15D]{Constraining the Circumbinary Disk Tilt in the KH 15D system}

\author[Poon, Zanazzi, \& Zhu]{
Michael Poon,$^{1,2}$\thanks{michaelkm.poon@mail.utoronto.ca (MP)}
J.~J. Zanazzi,$^{1}$\thanks{jzanazzi@cita.utoronto.ca (JJZ)}
and Wei Zhu,$^{1,3}$\thanks{weizhu@cita.utoronto.ca (WZ)}
\\
$^{1}$Canadian Institute for Theoretical Astrophysics, University of Toronto, 60 St. George Street, Toronto, Ontario, M5S 1A7, Canada\\
$^{2}$Department of Astronomy and Astrophysics, University of Toronto, Toronto, ON M5S 3H4, Canada\\
$^{3}$Department of Astronomy, Tsinghua University, Beijing, 100084, China
}

\date{Accepted XXX. Received YYY; in original form ZZZ}

\pubyear{2020}

\begin{document}
\label{firstpage}
\pagerange{\pageref{firstpage}--\pageref{lastpage}}
\maketitle

\begin{abstract}

KH 15D is a system which consists of a young, eccentric binary, and a circumbinary disk which obscures the binary as the disk precesses. We develop a self-consistent model that provides a reasonable fit to the photometric variability that was observed in the KH 15D system over the past 60 years. Our model suggests that the circumbinary disk has an inner edge $r_{\rm in}\lesssim 1 \ {\rm au}$, an outer edge $r_{\rm out} \sim {\rm a \ few \ au}$, and that the disk is misaligned relative to the stellar binary by $\sim$5-16 degrees, with the inner edge more inclined than the outer edge.  The difference between the inclinations (warp) and longitude of ascending nodes (twist) at the inner and outer edges of the disk are of order $\sim$10 degrees and $\sim$15 degrees, respectively. We also provide constraints on other properties of the disk, such as the precession period and surface density profile. Our work demonstrates the power of photometric data in constraining the physical properties of planet-forming circumbinary disks.

\end{abstract}

\begin{keywords} 
protoplanetary discs -- planets and satellites: formation -- stars: individual (KH 15D) -- binaries: spectroscopic -- techniques: photometric
\end{keywords}


\section{Introduction}
\label{sec:Intro}

Our understanding of tilts within planet-forming circumbinary systems has undergone drastic changes within the past decade.  Originally, the basic picture was quite simple: a circumbinary disk should always be observed to be aligned with the orbital plane of the binary.  Even though simulations of turbulent molecular clouds found circumbinary disks frequently formed misaligned with the  orbital plane of the binary \citep[e.g.][]{Bate(2002),Bate(2012),Bate(2018)}, viscous disk-warping torques were showed to damp the disk-binary inclination over timescales much shorter than typical protoplanetary disk lifetimes \citep{FoucartLai(2013),Foucart2014}.  Most inclination constraints on protoplanetary \citep[e.g.][]{Andrews(2010),Rosenfeld(2012),Czekala(2015),Czekala(2016),Ruiz-Rodriguez(2019)} and debris \citep[e.g.][]{Kennedy(2012b)} disks confirm this basic picture, finding alignment of the disk with the orbital plane of the binary to within a few degrees.

However, after the detection of a few highly inclined circumbinary disks \citep[e.g.][]{Kennedy(2012a),Marino2015,Brinch(2016),Czekala(2017)}, it became clear not all circumbinary disks align rapidly.  Motivated by these detections of highly-inclined disks, the theoretical community found that when a circumbinary disk orbits an \textit{eccentric} binary, the disk-binary inclination can grow under certain circumstances, evolving eventually to $90^\circ$ (polar alignment; \citealt{Aly(2015),MartinLubow(2017),Zanazzi2018}).  Additional inclined circumbinary disks orbiting eccentric binaries were discovered soon thereafter, such as HD 98800 \citep{Kennedy2019}, and AB Aurigae \citep{Poblete(2020)}.  Recently, \cite{Czekala2019} showed circumbinary disks had larger inclinations when orbiting binaries with higher eccentricities, further supporting the operation of this mechanism in circumbinary disk systems.

The alignment process itself was also shown to be non-trivial, with the disk itself occasionally breaking in the process.  Early on, the disk was expected to remain nearly flat, due to the resonant propagation of bending waves across the disk \citep{PapaloizouLin(1995),LubowOgilvie(2000)}.  Later hydrodynamical simulations found that the disk under some circumstances may break, with different disk annuli becoming highly misaligned with one another, due to strong differential nodal precession induced by the torque from the binary \citep[e.g.][]{Nixon(2013),Facchini(2013)}.  Numerous broken protoplanetary disks orbiting two binary stars have subsequently been found, including HD 142527 \citep{Marino2015,Price(2018)} and GW Ori \citep{Bi2020,Kraus(2020)}.

The inclinations of detected circumbinary planets, in contrast, remain broadly consistent with formation in nearly-aligned circumbinary disks.  After the detection of a few dozen circumbinary planets (see \citealt{WelshOrosz(2018),DoyleDeeg(2018)} for recent reviews), the inclinations within the circumbinary planet population are consistent with alignment to within $\sim 4^\circ$ \citep{MartinTriaud(2014),Armstrong(2014),Li(2016)}.  However, highly-misaligned circumbinary planets are physically allowed, because the inclined orbit has been shown to be long-term stable once a planet forms in the polar-aligned circumbinary disk.  \citep{DoolinBlundell(2011),GiupponeCuello(2019),Chen(2019),Chen(2020)}.  New detection methods may detect polar-aligned circumbinary planets in the future \citep{ZhangFabrycky(2019)}.

While a large number of systems now have constraints on mutual inclinations between the disk and the binary orbital planes, there remain few constraints on twists and warps within the circumbinary disk itself.  This is because the methods used to constrain disk inclinations are not sensitive to the small misalignments within the disk.  Gaseous protoplanetary disk inclinations are constrained via the orbital motion of the disk gas through the Doppler shift of emission lines \citep[e.g.][]{Facchini(2018),Price(2018)}. Debris disk inclinations are constrained by the orientation of the disk implied by its continuum emission \citep[e.g.][]{Kennedy(2012a),Kennedy(2012b)}.

A rare example of a circumbinary disk\footnote{
In this work, we use the terminology disk, rather than ring, to describe the object (likely) extending only to a few au around the KH 15D stellar binary.  Although this runs counter to more traditional ideas of what a protoplanetary disk is within the planet-formation community, which has envisioned a disk as a gaseous object orbiting a young stellar object out to tens or hundreds of au, recent observations have detected more compact protoplanetary disks.  \cite{Pegues+(2021)} found the CO emission around the young M-dwarf FP Tau extended to $\sim$4-8 au, while \cite{FrancisvanderMarel(2020)} resolved the size of the inner disks in a number of transition disk systems to lie near or within $\sim$1-10 au.
}
system where photometric constraints exist is Kearns-Herbst 15D (KH 15D) \citep{Kearns1998}.  KH 15D is a system with a highly-unusual light curve, which exhibited dips by up to 5 magnitudes. The morphology of the dipping behavior changed over decade-long timescales, but displayed periodicity over short 48 day timescales.  The complex light curve of this system is generally believed to be due to a circumbinary disk and a binary star, with some of the dips caused by the optically thick, precessing disk slowly and obscuring the orbital plane of the stellar binary \citep{Winn(2004),Winn2006,Chiang2004,Capelo(2012)}.  Although much work has gone into understanding KH 15D, no work has attempted to provide quantitative constraints on the properties of the warped disk based on the photometric data.

In this work, we combine the spectroscopic and photometric data to constrain the properties of the circumbinary disk KH 15D.  With recent data up to 2018 from \citet{Aronow2018} and \citet{GarciaSoto2020}, we improve the \cite{Winn2006} model to fit all photometric data since 1955. Our results are particularly exciting, as our fit nearly encompasses the full transit of the circumbinary disk of KH 15D.  Section~\ref{sec:model} extends the \cite{Winn2006} model to fit the {light curve of the system, over the more than 60 year duration the system was observed.}  Section~\ref{sec:Secular} develops a dynamical model to constrain the circumbinary disk properties implied by the photometric constraints.  Section~\ref{sec:Discuss} discusses the theoretical implications of our work, improvements which can be made to our model, and our model predictions which can be tested with future observations.  Section~\ref{sec:Conc} summarizes the conclusions of our work.

\section{Modelling The Light curve Of KH 15D}
\label{sec:model}

\subsection{Photometric \& Radial Velocity Observations}
\label{sec:model_obs}

We use radial velocity observations to constrain the orbit of the stellar binary and photometric data to constrain the geometry of the optically thick, precessing disk. We begin with a brief description of the radial velocity and photometric data used in this work.

We use the Radial Velocity (RV) measurements gathered by \citet{Hamilton(2003)} and \citet{Johnson2004}, and because one of the stars is occulted by the disk when the RV measurements are taken, this is effectively a single-lined spectroscopic binary.  As in \cite{Winn2006}, we only use RV measurements gathered when the system flux is 90\% or greater than its mean out-of-occultation flux, because the Rossiter-McLaughlin effect \citep{Rossiter(1924),McLaughlin(1924)} leads to systematic errors as the stellar companion is occulted by the disk\footnote{We note that the Rossiter-McLaughlin effect is only important in this system, if the KH 15D disk edges are sharp, not diffuse.}.  This gives 12 RV measurements to aid in constraining the orbit of the binary.

For the photometry, we use the tabulated data from \citet{Winn2006}, \citet{Aronow2018}, and \citet{GarciaSoto2020}.
Details about the photometric observations can be found in these references, and we only provide a brief summary here.  These catalogues include data from photographic plates from the 1950s to 1985 \citep{JohnsonWinn(2004),Maffei2005}, as well as observations using Charge-Coupled Devices (CCDs) since 1995 \citep[e.g.][]{2005Hamilton,Capelo(2012),Aronow2018,GarciaSoto2020}.  No observations were known to be taken between 1985-1994. All photometric observations have been transformed into the standard Cousin $I$-band measurements. We bin the original data-set (6241 points) into 2813 data points in order to reduce the amount of time needed for the photometric model computation. The uncertainties of all photometric measurements are re-scaled up by a factor of two, to allow for a model fit which gives a reduced $\chi^2$ close to unity.

\subsection{Previous models for KH 15D}
\label{sec:model_W06}

So far four models have been proposed to explain the photometric variation in the KH 15D system. The phenomenological model by \cite{Winn(2004),Winn2006}  approximates the leading edge of the disk as an infinitely long and optically thick screen, which occults the two stars as the screen moves across the orbit of the binary.  Motivated by dynamics, \citet{Chiang2004} treat the KH 15D disk as a warped disk with finite optical depth, and model the photometric variations by a disk precessing into and out of the line-of-sight of the observer.  \citet{SilviaAgol(2008)} developed their model based on the model of \cite{Winn2006}, but introduced more disk-related physics, such as the finite optical depth, curvature near the edge, and forward-scattering of starlight from the dust in the disk (which was parameterized as ``halos'' in the \citealt{Winn2006} model). The fourth model is that of \cite{GarciaSoto2020}, who extended the \cite{Winn2006} model to include a trailing, as well as leading, edge. We choose to build our model based on \cite{Winn2006}, because it allows us to remain agnostic about the detailed physics of the disk itself, while still accurately fitting the light curve of KH 15D.  We review the \cite{Winn2006} model within this subsection.

\begin{figure*}
    \centering
    \includegraphics[width=\linewidth]{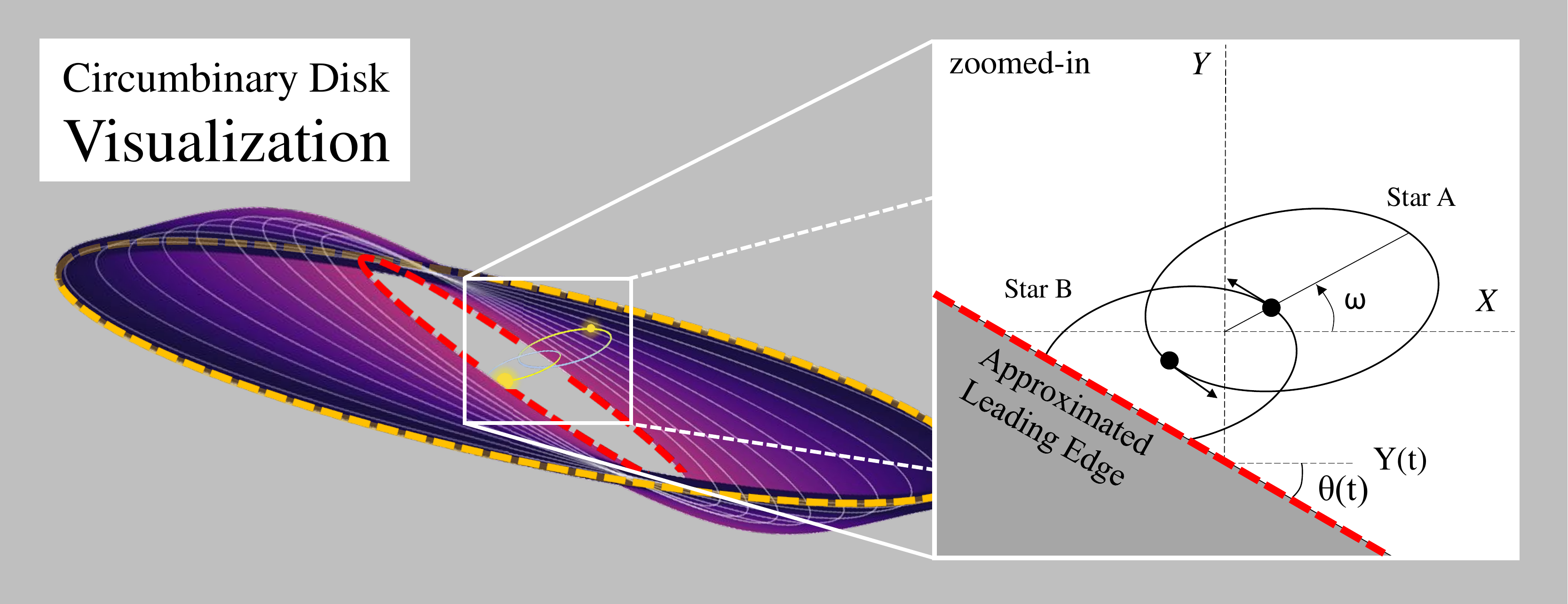}
    \caption{The purple ellipses display a potential shape for the circumbinary disk of KH 15D. No assumption about the inclination between the innermost edge (red dashed curve), outermost edge (yellow dashed curve) or binary plane (centre yellow ellipses) are made while modelling the occultations of KH 15D. We note that our model also allows for the leading edge to be the outermost truncation radius of the disk, with the trailing edge as the innermost truncation radius of the disk.  The zoomed-in inset diagram displays how the circumbinary disk geometry occults the binary of KH 15D.  The inner or outer truncation radius of the disk slowly covers the orbital plane of the binary, as the disk precesses around the orbital angular momentum axis of the binary.  We approximate the inner and outer disk edges as straight edges as the binary is occulted.
    }
    \label{fig:model1}
\end{figure*}

Figure~\ref{fig:model1} illustrates the physical motivation behind the \cite{Winn2006} model.  A single ``leading'' edge (red dashed line) slowly advances over the orbital plane of the binary, which approximates the inner or outer truncation radius of the disk slowly covering both stars as the disk is precessing around the binary.  For data taken before 2005, it is reasonable to neglect the outer disk truncation radius (the ``trailing'' edge, as marked by the yellow dashed curve).  To model how quickly the leading edge advanced across the orbit of the binary, \cite{Winn2006} used the latest date when star B was still visible ($t_4$), and the latest date when the orbit of star A was visible ($t_5$, see Fig.~\ref{fig:tdefined_combined} left panel), as free parameters in their model.  In addition, not only was the angle the leading edge made with the $X$-axis of the observer allowed to vary, but it was also allowed to change at a constant rate, controlled by two free parameters $\theta_L(t_4)$ and $\dot \theta_L$.  The  light from the binary was modelled with 7 parameters, with the luminosity from star A (B) denoted by $L_{\rm A}$ ($L_{\rm B}$), the background light when the disk fully occults the binary by $L_0$ , with the parameters $\{\epsilon_1, \epsilon_2, \xi_1, \xi_2\}$ parameterizing the light emitted by halos surrounding stars A and B.   Specifically, the 1D brightness distribution from star $i = {\rm A}, {\rm B}$ was taken to be
\be
B_i (v) = 
\left\{ \begin{array}{ll}
(\epsilon_1/\xi_1) \exp \left[ (v + 1)/\xi_1 \right] & v \le -1 \\
(\epsilon_1/\xi_1) + B_{\star i}(v) & -1 < v < 1 \\
(\epsilon_2/\xi_2) \exp \left[ -(v - 1)/\xi_2 \right] & v \ge 1
\end{array} \right. ,
\label{eq:B(v)}
\ee
where $B_{\star i}$ is the 1D brightness distribution of star $i$, assuming a linear limb-darkening model:
\be
B_{\star i}(v) = 2 I_i \sqrt{1 - v^2} \left[ 1 - u \left( 1 - \frac{\pi}{4} \sqrt{1 - v^2} \right) \right].
\ee
Here, $u = 0.65$ is the limb-darkening coefficient for both stars, and $I_i$ is the reference intensity of star $i$.  Letting $y_{L,i}$ be the distance of the lead edge from star $i$, and $v_{L,i} = y_{L,i}/R_i$, then the flux from star $i$ is
\be
F_{L,i} = \int_{v_{L,i}}^\infty B_i (v) \der v,
\ee
with the total flux $F = F_{L,{\rm A}} + F_{L,{\rm B}}$.  Physically, each ``halo'' parameterizes forward-scattering of starlight by dust in the disk \citep{Winn2006,SilviaAgol(2008)}.  The mass and radius for star A were taken to be $M_{\rm A} = 0.6 \, \Msun$ and $R_{\rm A} = 1.3 \, \Rsun$, while the ratios between the masses and radii of the two stars are $M_{\rm B}/M_{\rm A} = 1.2$ and $R_{\rm B}/R_{\rm A}=1.05$, respectively.  The orbit of the binary is described by standard orbital parameters used to model RV data, with an orbital period $P$, eccentricity $e$, inclination $I$, longitude of pericenter $\omega$, time of pericenter passage $T_p$, and line-of-sight velocity $\gamma$ (see e.g. \citealt{Fulton2018} for details).  The Cartesian coordinate system in the sky-projected reference plane of the observer $(X,Y)$ is chosen so the $X$-axis lies along the line of nodes (so $\Omega = 0$).

The best fit model parameters for the KH 15D system was then calculated by minimizing \citep{Winn(2004),Winn2006}
\begin{align}
\chi^2 &= \sum_{j=1}^{N_F} \left( \frac{F_j - F_{O,j}}{\sigma_{F,j}} \right)^2 + \lambda \sum_{j=1}^{N_V} \left( \frac{V_j - V_{O,j}}{\sigma_{V,j}} \right)^2
\nonumber \\
&\equiv \chi^2_{\rm phot} + \lambda \chi^2_{\rm RV},
\label{eq:chi2_W06}
\end{align}
where $\chi^2_{\rm phot}$ is the $\chi^2$ of the photometry model alone, $\chi^2_{\rm RV}$ is the $\chi^2$ metric of the modelled orbit of the binary in relation to the RV data (see \citealt{Fulton2018} for details), and for a quantity $X$, $X_j$ denotes the model prediction at point $j$, $X_{O,j}$ denotes the observed value of $X$ at $j$, while $\sigma_{X,j}$ denotes the uncertainty of $X_{O,j}$ at $j$.  The parameter $\lambda = 50$ was chosen to increase the importance of the RV model relative to that for the photometry, because the model constraining the orbit of the binary (a Keplerian orbit) is much more certain than the model describing the light curve of the binary (occulted by a precessing disk).

Because the screen advances in the positive vertical direction at a constant rate, an equivalent way of parameterizing the ascent of the screen are through where the screen intersects the $Y$-axis at the orbital contact time $t_4$, which we will denote by $Y_L(t_4)$ (see Figs.~\ref{fig:model1} \&~\ref{fig:tdefined_combined}), and the rate of change in the $Y$-direction, $\dot Y_L$.  \cite{Winn2006} choose $t_4$ and $t_5$ because of its tighter connections with observations ($t_4, t_5$ denote changes in the light curve of KH 15D).  When extending the \cite{Winn2006} model, we will also primarily refer to orbital contact times to parameterize the advance of the screen across the orbit of the binary (Fig.~\ref{fig:tdefined_combined}), but also frequently refer to $Y_L$ and $\dot Y_L$ as well.

\subsection{Our model for the KH 15D System}
\label{sec:model_our}

\begin{figure*}
    \centering
    \includegraphics[width=\linewidth]{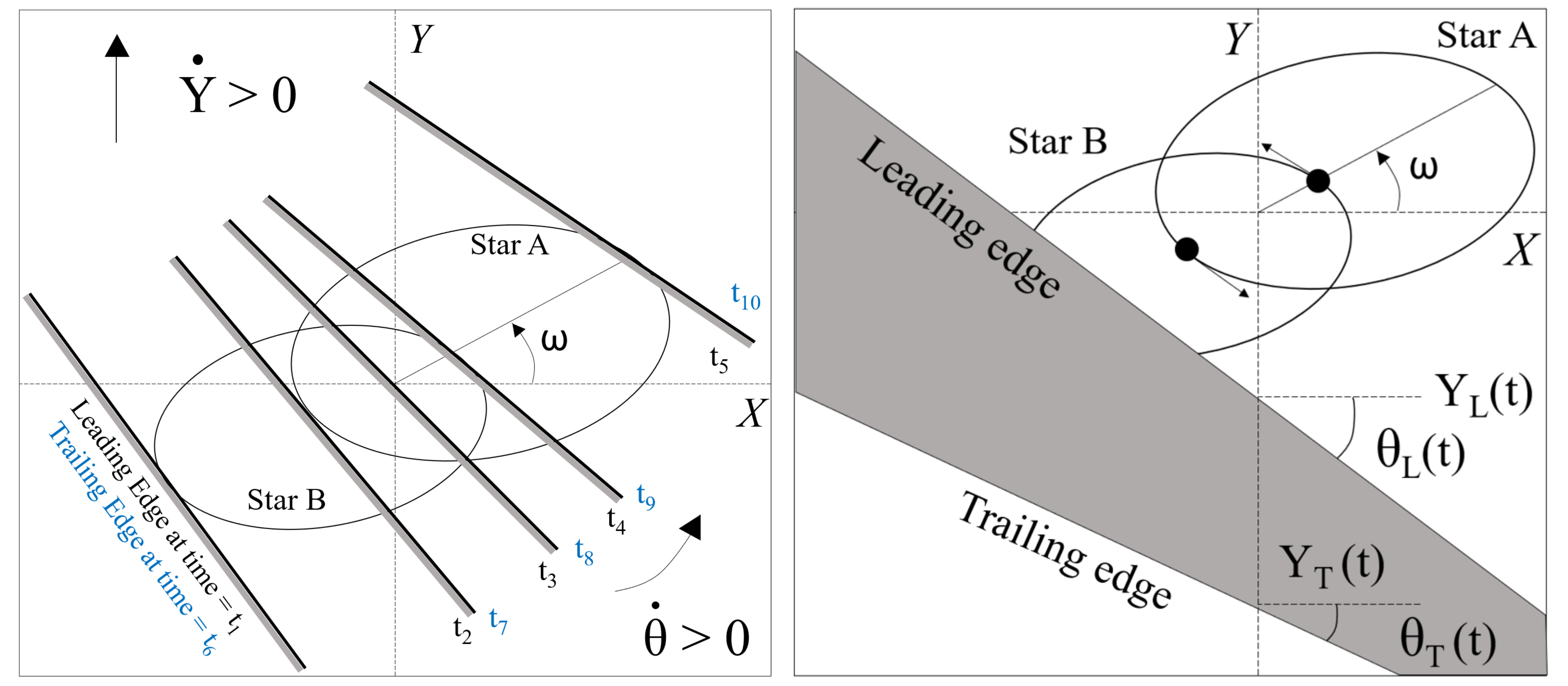}
    \caption{\textit{Left Panel}:  Definitions of orbital contact times.  The leading (trailing) edge has contact times $t_1$ to $t_5$ ($t_6$ to $t_{10}$).  All contact times denote when the leading or trailing edge lies tangent to the orbit of either star A or B, with the exception of $t_3$ and $t_6$, which denote when the leading or trailing edges intersect the center of mass of the binary.  \textit{Right Panel}: The definitions of quantities related to our model of the disk occulting the binary of KH 15D, which we model as an opaque screen bounded by two infinitely-long, straight edges on both sides.  The leading (trailing) edge is parameterized by its intersection with the $Y$-axis, $Y_L$ ($Y_T$), and the angle between the edge and the $X$-axis, $\theta_L$ ($\theta_T$).  Our model allows for $Y_L$, $\theta_L$, $Y_T$, and $\theta_T$ to evolve (linearly) with time.  We emphasize our model makes no assumption on the underlying geometry of the disk occulting the binary of KH 15D.
    }
    \label{fig:tdefined_combined}
\end{figure*}

Our photometric model builds off of \cite{Winn2006}, and seeks to fit the light curve of KH 15D from 1955-2018 with minimal modifications (smallest number of additional parameters to describe the trailing edge, in relation to the leading edge).  After 2012, the trailing edge started to uncover star B, due to the other (inner or outer) truncation radius of the disk of KH 15D precessing over the binary orbit with respect to the line-of-sight of the observer (see Fig.~\ref{fig:model1}).  The simplest extension is to include an additional trailing edge in the modelling (denoted by subscript $T$), which lags in position behind the leading edge (with subscript $L$), which intersects the $Y$-axis at a location $Y_T(t)$ (see Fig.~\ref{fig:tdefined_combined}).  This trailing edge also introduces 5 new orbital contact times as the edge crosses the orbit of the binary: $t_6,t_7,t_8,t_9,t_{10}$ (see Fig.~\ref{fig:tdefined_combined} for illustration). Assuming $\theta_L = \theta_T$ and $\dot Y_L = \dot Y_T$, the previous 1-edge semi-infinite sheet becomes a 2-edge thin rectangular sheet of constant width, which is infinite along its length. \cite{GarciaSoto2020} assumed this for their light curve model, and neatly fit CCD photometry from 1995 and onwards. However, because this fit does not match the light curve data prior to 1995 (fit not shown here or in \citealt{GarciaSoto2020}), further modifications are needed to the \citet{Winn2006} and \citet{GarciaSoto2020} model. 

\begin{table}
\caption{
    Definitions of Model Parameters.
}
\label{tab:description}
\begin{threeparttable}
\begin{tabular}{ll}
\hline\hline
    Free Parameter & Description \\ \hline
    $P$ & Orbital period \\
    $e$ & Orbital eccentricity \\
    $I$ & Inclination of orbital plane \\
    $\omega$ & Argument of pericenter \\
    $T_p$ & Time of periapsis passage \\
    $L_B/L_A$ & Luminosity of star B relative to star A \\
    $\epsilon_1$ & Fractional flux of stellar halo\tnote{1} \\
    $\epsilon_2$ & Fractional flux of stellar halo\tnote{2} \\
    $\xi_1$ & Exponential scale factor of stellar halo\tnote{1} \\
    $\xi_2$ & Exponential scale factor of stellar halo\tnote{2} \\
    $t_3$ & Third orbital contact time\tnote{3}  \\
    $t_5$ & Fifth orbital contact time\tnote{3}  \\
    $t_6$ & Sixth orbital contact time\tnote{3}  \\
    $\theta_L(t_3)$ & Angle between x-axis and leading edge at $t=t_3$ \\
    $\theta_T(t_3)$ & Angle between x-axis and trailing edge at $t=t_3$ \\
    $\dot{\theta}_{L1}$ & Rotation rate of leading edge when $t<t_3$\\
    $\dot{\theta}_{L2}$ & Rotation rate of leading edge when $t>t_3$\\
    $\dot{\theta}_{T}$ & Rotation rate of trailing edge \\
\hline
\end{tabular}
\begin{tablenotes}
\item[1] In the direction the leading edge approaches the star
\item[2] In the direction the leading edge travels beyond the star
\item[3] Defined in \autoref{fig:tdefined_combined}
\end{tablenotes}
\end{threeparttable}
\end{table}

Through much experimentation, we found the following set of additions to the \cite{Winn2006} model that let us fit the 60+ year light curve. The connection of these additions with a warped disk driven into precession by an eccentric binary will be made clear in the following section.
\begin{itemize}
\item We let the leading and trailing edges have different angles ($\theta_L[t] \ne \theta_T[t]$, see Fig.~\ref{fig:tdefined_combined}).
\item We let each edge linearly evolve in time independently ($\dot{\theta}_L[t] \neq \dot{\theta}_T[t]$).
\item Parameterize $\dot \theta_L$ by two constant, piecewise rates in time: $\dot \theta_L(t) = \dot \theta_{L1}$ when $t < t_3$, and $\dot \theta_L(t) = \dot \theta_{L2}$ when $t > t_3$.  We keep $\dot \theta_T(t) = {\rm constant}$ as a single parameter.  Because we make the leading edge symmetric about $t_3$, we fit for the times $\{t_3,t_5\}$ in our MCMC model to constrain $Y_L(t_3)$ and $\dot Y_L$, rather than $\{t_4,t_5\}$ as in \cite{Winn2006}.
\item Let the width of the screen change over time ($\dot Y_L \ne \dot Y_T$), but keep both rates $\dot Y_L$ and $\dot Y_T$ constant with time. 
\item Prescribe the rate of ascent of the trailing edge in relation to the rate of ascent of the leading edge.  Specifically, we take $\dot{Y}_T=\alpha\dot{Y}_L$ for $\alpha=0.1,0.3,0.5,2.0,3.0,10.0$. We also experimented with letting $\dot{Y}_T$ be a free parameter (fitting for the contact times \{$t_6,t_7$\}), and found these fits gave $\dot{Y}_T \approx \dot{Y}_L$, but the MCMC did not always converge.  We choose this parameterization to make sure the other model parameters are well-determined.
\end{itemize}
We further simplify the Winn model by analytically solving for $L_A$ and $\gamma$ with respect to the rest of the parameters, since they are constant shifts to the photometric and radial velocity models, respectively. This reduces the number of free parameters by 2.  For reference, we display each model parameter and its definition in Table~\ref{tab:description}.

To calculate the flux from the KH 15D system, we simply add the flux from stellar light emitted exterior to the trailing edge, to that emitted exterior to the leading edge.  In more detail, letting $y_{T,i}$ be the distance of the trailing edge from star $i$, with $v_{T,i} = y_{T,i}/R_i$, star $i$ emits the flux
\be
F_{T,i} = \int_{-\infty}^{v_{T,i}} B_i(v) \der v
\ee
exterior to the trailing edge, giving the total flux $F = F_{L,{\rm A}} + F_{L,{\rm B}} + F_{T,{\rm A}} + F_{T,{\rm B}}$.  We neglect the intersection between the leading and trailing edges in the flux calculation, because this intersection occurs far from the orbit of the binary.

For our radial velocity model, we follow equations (2) and (3) from Section 2.1 of \cite{Fulton2018}. To optimize the model parameters, we use a Python-implemented Markov chain Monte Carlo (MCMC) package \texttt{emcee} by \cite{Foreman-Mackey2018}. We use the same $\chi^2$ statistic as in equation~\eqref{eq:chi2_W06}.

\begin{table}
\centering
\caption{
    Model fits to photometric and radial velocity data for the KH 15D system, taking $\alpha=0.5$.  Orbital parameters $\{P,e,I,\omega,T_p\}$ are constrained using photometry and radial velocity data, while the other parameters are constrained using photometry alone.
    }
\label{tab:best_fit_new}
\begin{threeparttable}
\begin{tabular}{lr}
\hline
    Parameter & Our Fit \\ \hline
    $P$ [days] & $48.3777 \substack{+0.0002 \\ -0.0002}$\\
    $e$ & $0.5784 \substack{+0.0009 \\ -0.0009}$\\
    $I$ [deg] & $91.001 \substack{+0.002 \\ -0.001}$\\
    $\omega$ [deg] & $11.80 \substack{+0.06 \\ -0.06}$\\
    $T_p$ [JD] - 2,452,350 & $4.18 \substack{+0.01 \\ -0.02}$\\
    $L_B/L_A$ & $1.65 \substack{+0.01 \\ -0.01}$ \\
    $\epsilon_1$ & $0.0436 \substack{+0.0006 \\ -0.0006}$ \\
    $\epsilon_2$ & $0.0591 \substack{+0.0008 \\ -0.0008}$ \\
    $\xi_1$ & $1.53 \substack{+0.03 \\ -0.03}$\\
    $\xi_2$ & $2.86 \substack{+0.03 \\ -0.03}$\\
    $t_3$ & $1992.68 \substack{+0.05 \\ -0.05}$\\
    $t_5$ & $2007.95 \substack{+0.01 \\ -0.01}$\\
    $t_6$ & $2013.57 \substack{+0.03 \\ -0.03}$\\
    $\theta_L(t_3)$ [deg] & $-16.0 \substack{+0.2 \\ -0.2}$\\
    $\theta_T(t_3)$ [deg] & $-5.3 \substack{+0.2 \\ -0.2}$\\
    $\dot{\theta}_{L1}$ [rad/year] & $0.0077 \substack{+0.0002 \\ -0.0002}$\\
    $\dot{\theta}_{L2}$ [rad/year] & $0.0033 \substack{+0.0001 \\ -0.0001}$\\
    $\dot{\theta}_T$ [rad/year] & $-0.0006 \substack{+0.0001 \\ -0.0002}$\\ \hline
    $\chi^2_{\rm phot}$ & 13325 \\
    $\chi^2_{\rm RV}$ & 13 \\
    $\text{Reduced } \chi^2$ & 1.36 \\ \hline
    $t_1$ & $1972.9 \pm 0.2 $\tnote{1}\\
    $t_2$ & $1987.00 \pm 0.01 $\tnote{1}\\
    $t_4$ & $1996.8 \pm 0.1 $\tnote{1}\\
    $t_7$ & $2020.8 \pm 0.1 $\tnote{1}\\
    $t_8$ & $2024.95 \pm 0.01 $\tnote{1}\\
    $t_9$ & $2028.6 \pm 0.1 $\tnote{1}\\
    $t_{10}$ & $2041.0 \pm 0.5 $\tnote{1}\\ \hline
    $Y_T(t_3)$ [au] & -0.05903\tnote{2} \\
    $Y_L(t_6)$ [au] & 0.07642\tnote{2} \\
    $Y_T(t_6)$ [au] & -0.02082\tnote{2} \\
    $\dot Y_L(t_6)$ [au/year] & 0.003658\tnote{2} \\
    $\dot Y_T(t_6)$ [au/year] & 0.001829\tnote{2} \\  \hline
\end{tabular}
\begin{tablenotes}
\item[1] Predicted by the free parameters.
\item[2] Best-fit value, we do not calculate the errors implied by the $t_i$ measurements.
\end{tablenotes}
\end{threeparttable}
\end{table}

Preliminary tests find  the background light in the KH 15D system to be $L_0 \approx 0$, so we remove $L_0$ from our model parameters.  This is expected if $L_0$ is from forward scattering of the stars' light around the trailing edge of the disk \citep{SilviaAgol(2008)}, rather than the finite optical depth of the disk itself \citep{Chiang2004}, because forward scattering of stellar light around both screen edges is included in our model.  Our final model has 18 parameters (see \autoref{tab:description}), which we run for 20,000 steps with 36 walkers. Running the final model for each $\alpha$, we come to the following results: each MCMC converged except for $\alpha=10.0$, with only $\alpha=0.3,0.5$ producing reasonable-looking light curves. Model parameters for $\alpha=0.1,0.3,2.0,3.0$ are reported in Appendix~\ref{tab:best_fit_new2}. We highlight $\alpha=0.5$ as the best fit with parameters in \autoref{tab:best_fit_new}, and display corner plots of the posteriors in Appendix~\ref{fig:corner}. Because no stellar eclipses have been detected in the KH 15D light curve (only the disk-binary occultations), we fix the binary inclination to be $I \ge 91^\circ$ in our MCMC analysis, so the system is not perfectly edge-on.

\begin{figure*}
    \centering
    \includegraphics[width=\linewidth]{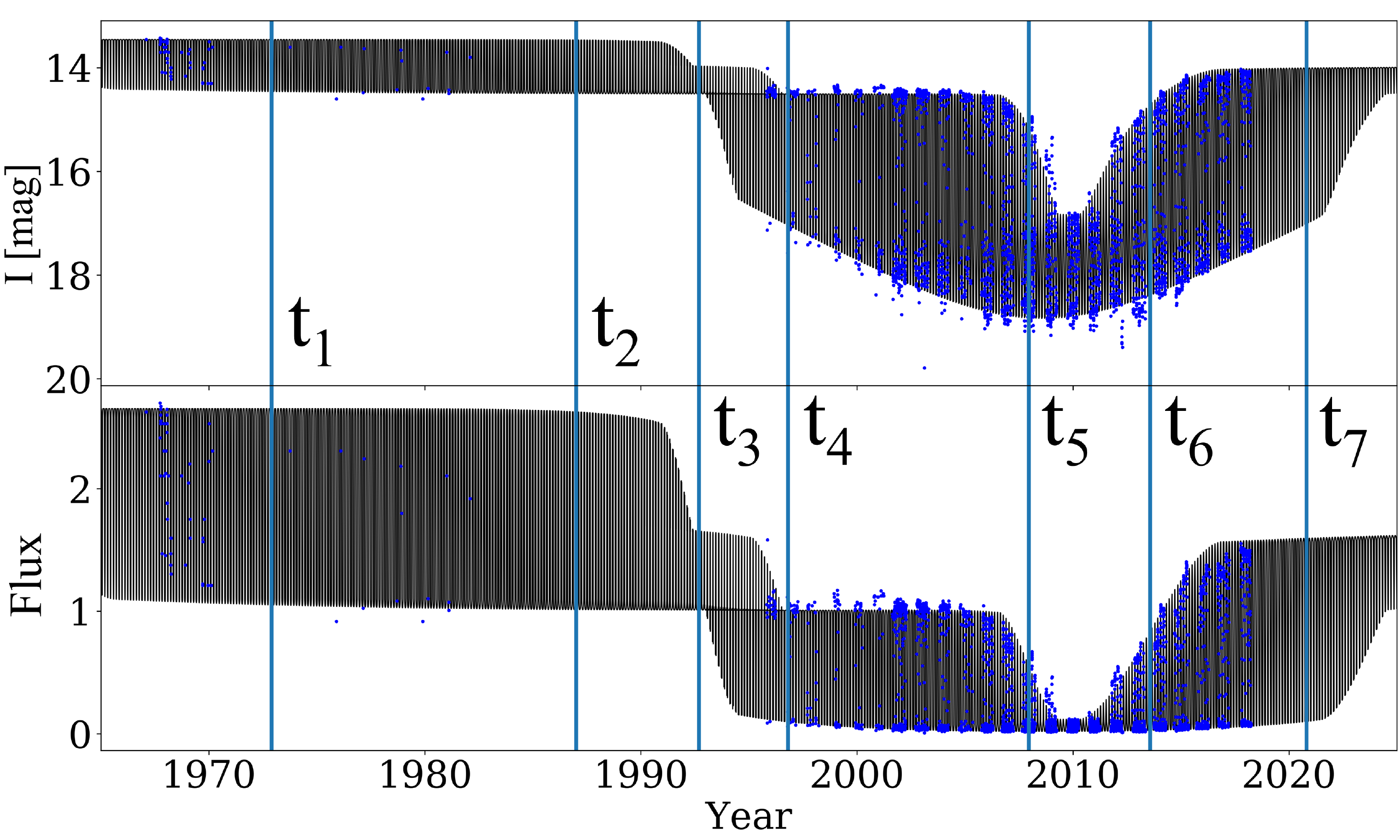}
    \caption{Light curve of KH 15D from 1965 to present, displaying the complex change in variability seen with time.  The observed light curve in $I$-band magnitude is shown in the upper panel, and the light curve after the normalization to the flux of star A is shown in the lower panel.  Blue points are photometry from \protect\citet{Aronow2018} and \protect\citet{GarciaSoto2020}, while the thin black line displays our photometric model fit (see Table~\ref{tab:best_fit_new} for parameter values).  Vertical cyan lines denote the orbital contact times $t_i$ indicated, where the leading or trailing edge of the screen (e.g. circumbinary disk, see Fig.~\ref{fig:model1}) hits a different portion of the binary orbit (see Fig.~\ref{fig:tdefined_combined} for definitions).  Our model does well in reproducing the KH 15D light curve variability over the length of time the system is observed.
    }
    \label{fig:unfolded}
\end{figure*}

Our fit for the entire light curve of KH 15D is displayed in Figure~\ref{fig:unfolded}.  Our model does a good job in describing both the maximum and minimum fluxes from KH 15D, which change with time.  As expected, the orbital contact times $t_i$ denote when the light curve of the system changes its morphology.  The gradual change of maximum/minimum flux around $t_i$ values is due to the halos around each star: for point-source stars occulted by a razor-thin opaque edge, the photometric model predicts almost discontinuous changes in light curve morphologies around $t_i$ values.

\begin{figure*}
    \centering
    \includegraphics[width=\linewidth]{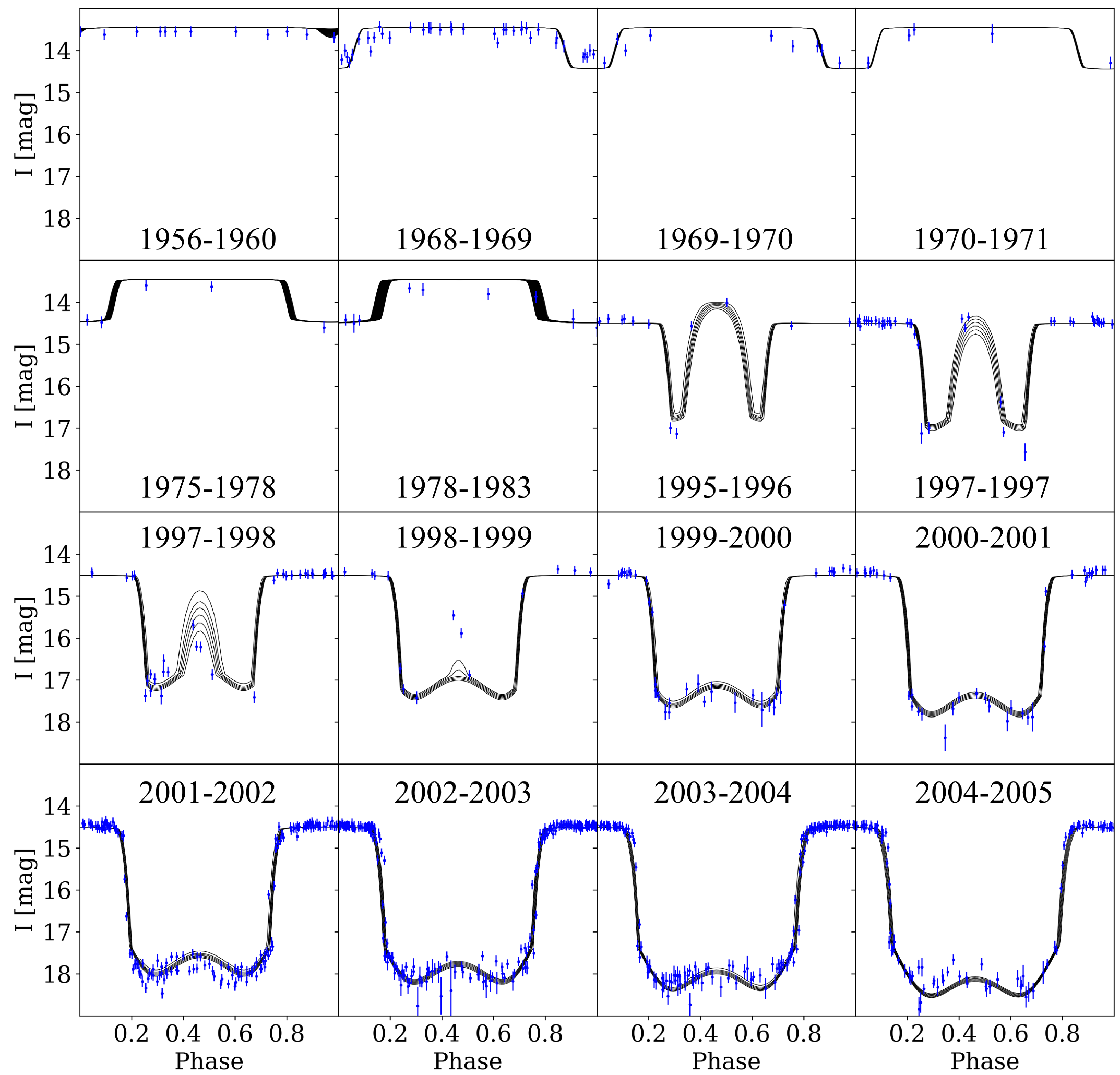}
    \caption{Data (blue points) and fitted model (black lines) displayed in Figure~\ref{fig:unfolded}, folded over the binary orbital period, prior to the year 2005.  The timespan over which the data and model are folded over is displayed in each figure.  Our model reproduces the changing morphology of the light curve of KH 15D well.
    }
    \label{fig:folded1}
\end{figure*}

\pagebreak

\begin{figure*}
    \centering
    \includegraphics[width=\linewidth]{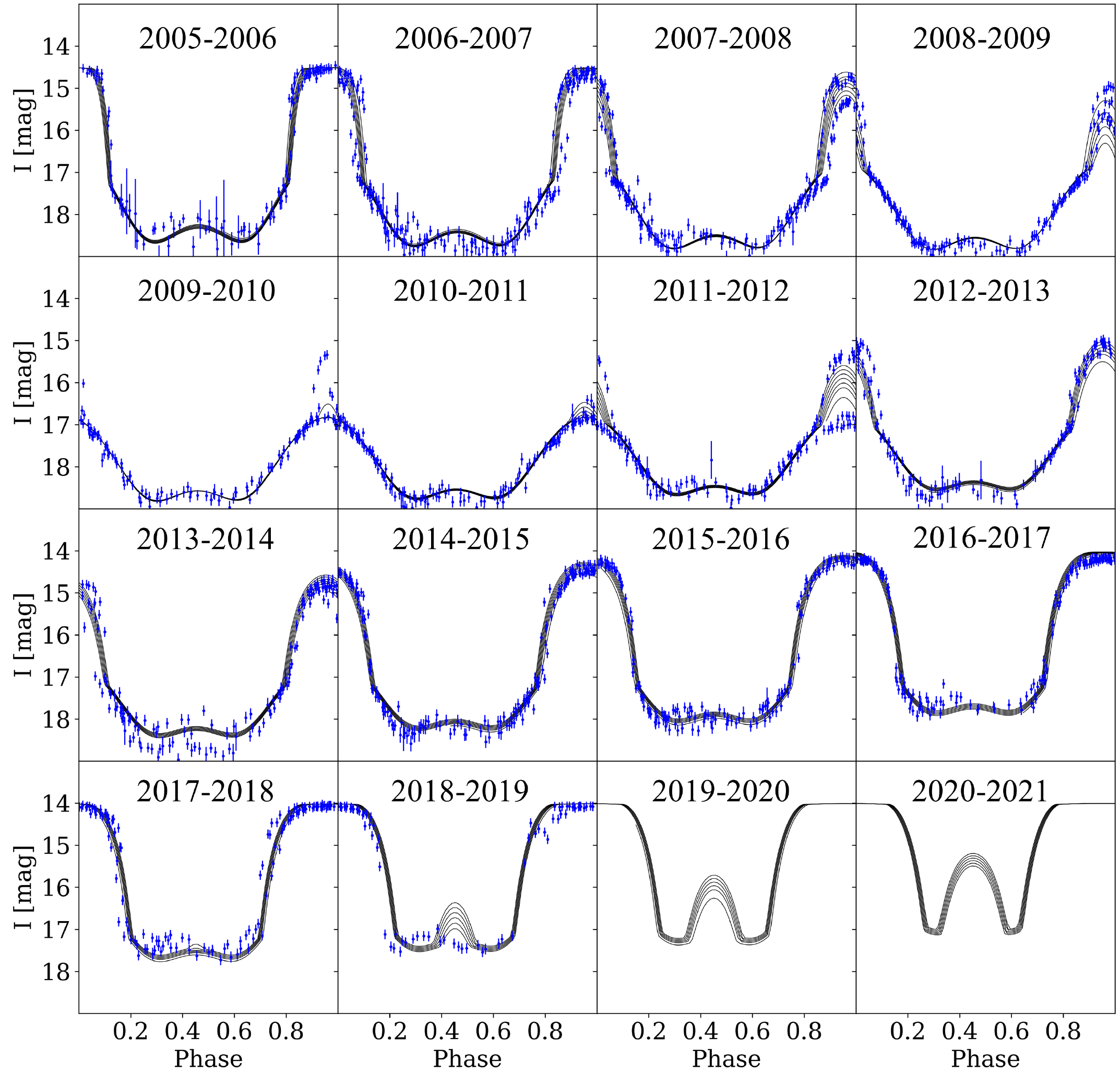}
    \caption{Same as Figure~\ref{fig:folded1}, except for data and model fits after the year 2005.  We also display model predictions for the years 2019 to 2021.}
    \label{fig:folded2}
\end{figure*}

\begin{table}
\centering
\caption{
    An excerpt from Table~\ref{tab:online_data}. For interested observers, the full version (available in machine-readable form) includes $I$ band magnitude predictions for the years 2000-2050 from our light curve model of KH 15D as shown in Figure~\ref{fig:unfolded}, Figure~\ref{fig:folded1} and Figure~\ref{fig:folded2}.
    }
\label{tab:online_data}
\begin{tabular}{ccc}
\hline
    Julian Date & Gregorian Date & KH 15D I-band magnitude\\ \hline
    2451545.0 & 2000.000 & 14.509\\
    2451546.0 & 2000.003 & 14.507\\
    2451547.0 & 2000.005 & 14.506\\
    ... & ... & ...\\
    2469807.0 & 2049.999 & 13.462\\  \hline
\end{tabular}
\end{table}

\pagebreak

Figures~\ref{fig:folded1}-\ref{fig:folded2} show the photometric model and data folded over the binary orbital period, which is comparable to Figure~12 in \cite{Winn2006}.  Again, we see our model does a good job at modelling changes in the light curve of KH 15D, with orbital contact times (see Table~\ref{tab:best_fit_new} for values) delineating morphology changes as one or both stars becomes occulted or revealed by an edge.  Examining the data from 2013-2014 and 2015-2016, the large scatter makes it seem unlikely that any (simple) model could provide an accurate fit to the observed light curve.  In addition, we do not remove any outliers (compare 1998-1999 panel in Fig.~\ref{fig:folded1} to Fig.~12 in \citealt{Winn2006}). An interesting feature occurs around 2010, when the egress is poorly fit (for all values of $\alpha$). This could be related to the clumpiness/transparency near the edges of the disk as discussed in \cite{GarciaSoto2020}, where the assumption of sharp edges breaks down.

\section{A Dynamical Model for the Disk of KH 15D}
\label{sec:Secular}

The previous section showed that in order for the \cite{Winn2006} model to fit the entire more than 60 year light curve of KH 15D, a number of modifications to this original model must be made.  In this section, we illustrate how these modifications are motivated by the dynamics of a warped disk, driven into precession around an eccentric binary.  In doing so, we will show that the disk orientation, warp, radial extent, and even surface density profile may be constrained by photometry alone.

\subsection{Model for a Precessing, Warped Circumbinary Disk Orbiting KH 15D}
\label{sec:Secular_model}

For the disk around KH 15D to coherently precess over its lifetime, internal forces within the disk must keep neighboring disk annuli nearly aligned with one another, otherwise differential nodal precession from the gravitational influence of the binary will disrupt and ``break'' the disk \citep[e.g.][]{LarwoodPapaloizou(1997),Facchini(2013),Nixon(2013),MartinLubow(2018)}.  When these internal torques are much stronger than the external torque on the disk from the binary, the disk behaves as a rigid body, coherently precessing about the orbital angular momentum axis of the binary \citep[e.g.][]{MartinLubow(2017),Smallwood(2019),Moody(2019)}.  To model the dynamical evolution of the disk, we will assume the disk behaves approximately like a rigid plate, treating the disk as a secondary whose mass is distributed between radii $r_L$ and $r_T$.

However, before we introduce our model for an extended disk, we discuss the dynamics of a test particle on a circular orbit (which we will refer to as a ring), driven into precession by the torque from the binary.  Many authors have shown the orbital angular momentum unit vector of the ring $\blr$ is driven into precession and nutation about either the orbital angular momentum unit vector of the binary $\bl$, or eccentricity vector of the binary ${\bm e}$ (vector in pericenter direction with magnitude $e$).  The dynamical evolution of the ring depends sensitively on the initial orientation of $\blr$ with respect to $\bl$ and $\ve$, as well as the magnitude of the eccentricity of the binary $e$.  To calculate the evolution of $\blr$ about $\bl$ and $\ve$, we adopt the formalism of \cite{Farago2010}, who calculated the secular evolution of a ring about a massive binary with an eccentric orbit, after expanding the Hamiltonian of the binary to leading order in $r/a$ (where $r$ is the semi-major axis of the test particle), and averaging over the mean motions of the test particle and the binary.  It was found the characteristic precession and nutation frequency of $\blr$ about the binary was given by (denoted by $\alpha$ in \citealt{Farago2010})
\be
\nu = \frac{3 \mu}{4 M_{\rm t}} \left( \frac{G M_{\rm t}}{a^3} \right)^{1/2} \left( \frac{a}{r} \right)^{7/2},
\label{eq:nu}
\ee
where $M_{\rm t} = M_{\rm A} + M_{\rm B}$ is the total mass of the binary, while $\mu = M_{\rm A} M_{\rm B}/M_{\rm t}$ is the reduced mass of the binary.


After calculating the evolution of a (circular) test particle $\blr$ vector about $\bl$ and $\ve$ using \cite{Farago2010}, we then translate the evolution of $\blr$ into the inclination of the test particle $\Ir$ and longitude of ascending node $\Omr$, in the frame where ${\bm {\hat z}} = \bl$ and the line of nodes points in the direction of $\ve$.  Because the orientation of the binary orbit in the reference frame of a distant observer is described by the orbital elements $\{a,e,\om,I,\Om\}$, the position of the ring in the frame of the observer is
\be
\left( \begin{array}{c} x \\ y \\ z \end{array} \right)_{\rm r,obs} = R_Z(\Omega) R_X(I) R_Z(\om) R_Z(\Omr) R_X(\Ir) \left( \begin{array}{c} x_{\rm r} \\ y_{\rm r} \\ 0 \end{array} \right),
\ee
where $(x_{\rm r},y_{\rm r}) = r (\cos \varphi, \sin \varphi)$ parameterizes the $(X,Y)$ coordinates of the ring in the frame where ${\bm {\hat z}} = {\bm {\hat l}}$, and $R_X[\beta]$ ($R_Z[\gamma]$) denote rotations along the $X$ ($Z$)-axis by angles $\beta$ ($\gamma$).  As in \cite{Winn2006}, we choose the reference plane of the observer so the $X$-axis points along the binary line-of-nodes (so $\Om = 0$).  Also, because our MCMC model highly favors a nearly edge-on orbit (Table~\ref{tab:best_fit_new}), we assume $I \simeq 90^\circ$ for simplicity for the rest of this section.  All other orbital parameters are taken as their most likely values from Table~\ref{tab:best_fit_new}.

\begin{figure}
    \centering
    \includegraphics[width=\linewidth]{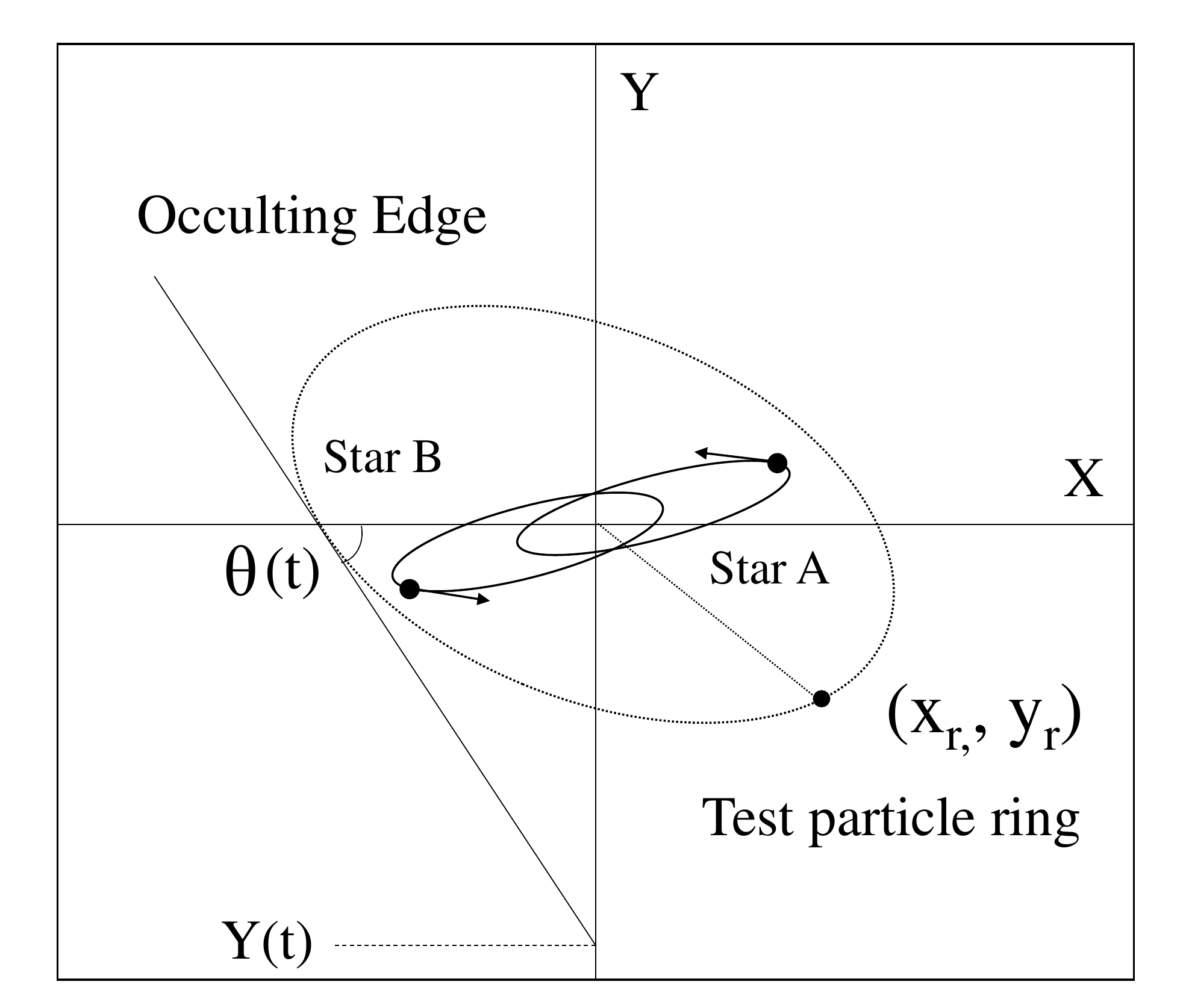}
    \caption{Our interpretation for the leading/trailing edges of the opaque screen in our photometry model.  The leading/trailing edges of the screen are from the inner or outer disk truncation radii.  The occulting disk edge is approximated by a straight line, drawn tangent to the intersection of the ring with the $X$-axis of the coordinate system.  The $\theta = \theta_k$ and $Y = Y_k$ values of the leading/trailing edge are defined similarly as in Figure~\ref{fig:tdefined_combined}.  Because the binary is nearly edge-on, the straight-line approximation is excellent.
    }
    \label{fig:1ringmodel_v2}
\end{figure}

To connect with a model for a disk occulting the binary of KH 15D, we approximate the inner and outer edges of the disk as two rings with different orbital elements $\{r_k, I_k, \Om_k\}$, with $k = L,T$ for the leading and trailing edges of the disk, respectively.  Although each ring has a different $r_k$, we assume the rings precess about $\bl$ with the same global disk precession frequency $\nu_{\rm d}$.

To connect the geometry of a disk occulting the binary of KH 15D with the edges of the light curve model in Section~\ref{sec:model}, we approximate an occulting ring by a line drawn tangent to the ring at the location where the ring intersects the $X$-axis of the system (see Fig.~\ref{fig:1ringmodel_v2}).  The angle between the tangent line and $X$-axis $\theta_k$, as well as the $Y$-intercept $Y_k$, of ring $k$ are then given by
\begin{align}
\theta_k \left[ I_k(t),\Omega_k(t) \right] &= \tan^{-1} \left[ \frac{\tan I_k}{\sin (\omega + \Omega_k)} \right],
\label{eq:theta} \\
Y_k\left[ I_k(t),\Omega_k(t) \right] &= -\frac{r_k \tan I_k}{\tan(\omega + \Omega_k)}.
\label{eq:Y}
\end{align}
A successful model of the circumbinary disk of KH 15D would give values for $\theta_k$ and $Y_k$ which match the MCMC fits for $\theta_L$, $\theta_T$, $Y_L$, and $Y_T$, from Table~\ref{tab:best_fit_new}.

\subsection{Estimates of Disk Properties from Model Fits}
\label{sec:Secular_est}

Before presenting an example warped disk geometry which matches the light curve model fit, we discuss how the warped disk geometry can be constrained by the MCMC fits of Section~\ref{sec:model_our}.  To do this, we simplify equations~\eqref{eq:theta}-\eqref{eq:Y}, and derive order-of-magnitude estimates for all disk quantities.  Because the  pericenter direction of the binary is nearly perpendicular to the observer ($\om \ll 1$), the disk annuli longitude of ascending nodes satisfy $\Om_k \approx \pi/2$ during transit.  The disk inclination is also nearly aligned with the orbital plane of the binary ($|I_k| \ll 1$).  Also, because the binary pericenter direction is nearly perpendicular to the observer, the inclination nutations should be near a local minimum \citep[e.g.][]{Farago2010,Zanazzi2018}, so $\dot I_k \approx 0$.  The nodal regression rate of the rings should be of order $\dot \Omega_k \approx -\nu_k$, where $\nu_k$ is the nearly constant nodal precession rate of ring $k$.  Defining $\dg \Omega_k \equiv \Omega_k - \pi/2$, and assuming $|\om|$, $|I_k|$, and $|\dg \Om_k| \ll 1$, equations~\eqref{eq:theta}-\eqref{eq:Y} can be shown to reduce to
\begin{align}
\theta_k &\approx I_k
\label{eq:theta_app} \\
\dot \theta_k &\approx -(\om + \dg \Om_k) I_k \nu_k,
\label{eq:dtheta_app} \\
Y_k &\approx r_k I_k (\om + \dg \Om_k),
\label{eq:Y_app} \\
\dot Y_k &\approx -r_k I_k \nu_k.
\label{eq:dY_app}
\end{align}
From this, we see the increase of $Y_L$ and $Y_T$ is primarily due to nodal regression from the rings.  The evolution of $\theta_L$ and $\theta_T$ is primarily due to the curvature of the ring, as it nodally precesses in front of the orbit of the binary (see \citealt{SilviaAgol(2008)} for further discussion).  Most interestingly, the MCMC constraints on $\theta_L$ and $\theta_T$ \textit{directly} translate to constraints on the ring inclinations $I_L$ and $I_T$.

Assuming the disk precesses rigidly ($\nu_L \approx \nu_T$), one can then constrain the disk radial extent.  Equation~\eqref{eq:dY_app} leads to
\be
\frac{r_L}{r_T} \approx \frac{\theta_T}{\theta_L} \frac{\dot Y_L}{\dot Y_T} = 0.63 \left( \frac{\theta_T}{-5^\circ} \right) \left( \frac{-16^\circ}{\theta_L} \right) \left( \frac{0.5}{\alpha} \right).
\ee
Because the values of $\alpha$ which fit the data are of order unity ($0.3 \lesssim \alpha \lesssim 1$), we can be confident that the leading edge of the screen occulting the binary of KH 15D is the disk inner truncation radius, while the trailing edge is the outer truncation radius ($r_L \lesssim r_T$).

Moreover, because $\dot \theta_k$, $Y_k$, and $\dot Y_k$ are all known, one can get unique solutions for $r_k$, $\dg \Omega_k$, and $\nu_k$.  Starting with $\nu_k$, equations~\eqref{eq:dtheta_app}-\eqref{eq:dY_app} can be re-arranged to give
\be
\nu_k \approx \left( \frac{\dot \theta_k \dot Y_k}{Y_k} \right)^{1/2}.
\label{eq:nuk_app}
\ee
Evaluating estimate~\eqref{eq:nuk_app} at $t = t_4$, we find $\nu_L \sim 0.013 \, {\rm yr}^{-1}$ and $\nu_T \sim 0.0073 \, {\rm yr}^{-1}$, which are consistent with one another within a factor of a few.  Similarly, equation~\eqref{eq:dY_app} can be solved for $r_k$:
\be
r_k \approx - \frac{1}{\theta_k} \left( \frac{Y_k \dot Y_k}{\dot \theta_k} \right)^{1/2},
\ee
which gives $r_L \sim 1.0 \, {\rm au}$ and $r_T \sim 4.6 \, {\rm au}$ at $t = t_4$ for our model.  Last, either equation~\eqref{eq:dtheta_app} or~\eqref{eq:Y_app} can be solved for $\dg \Omega_k$.

Although these disk parameter estimates are far from unique, they provide constraints on the properties of the circumbinary disk within the KH 15D system.  We can strongly conclude the disk-binary mutual inclination $I_{\rm KH \ 15D}$ in the KH 15D system lies in the range $5^\circ \lesssim I_{\rm KH \ 15D} \lesssim 16^\circ$, with the disk inner edge more highly inclined than the outer edge (because $I_L \gtrsim I_T$).  The leading edge of the opaque screen crossing the binary orbit is from the disk inner edge, which is located at a radius $r_L \lesssim 1 \, {\rm au}$, while the trailing outer disk edge is located at $r_T \sim {\rm few} \ {\rm au}$.

\subsection{Example Warped Disk which Matches Model Fits}
\label{sec:Secular_ex}

As we saw in the previous section, for a unique match to the phenomenological parameters $\{\theta_k,\dot \theta_k,Y_k,\dot Y_k\}$ to a precessing, inclined ring annulus, we require the ring parameters $\{r_k, I_k, \Omega_k, \nu_k\}$.  However, for a protoplanetary disk to exist over many dynamical times, it must precess rigidly ($\nu_L = \nu_T$), decreasing the number of free parameters in one ring.  Therefore, our dynamical model is over-determined by our phenomenological model.
To get accurate constraints on the warped disk itself using photometry, a light curve model must be developed whose free parameters are directly related to the warped disk properties (disk inclination, warp, twist, precession frequency, etc.), rather than indirectly through a phenomenological model.  The goal of this section is not to provide stringent constraints on the disk itself, but to present an example warped disk which gives gross light curve features consistent with the MCMC light curve fits.

\begin{table}
\centering
\begin{tabular}{lr}
\hline
    Parameter & Example Value \\ \hline
    $\nu_{\rm d}$ [${\rm yr}^{-1}$] & 0.01\\
    $r_L$ [AU] & 0.5\\
    $r_T$ [AU] & 2.0\\
    $I_L(t_3)$ [deg] & -13\\
    $I_T(t_3)$ [deg] & -6\\
    $\Omega_L(t_3)$ [deg] & 100\\
    $\Omega_T(t_3)$ [deg] & 115\\
\hline
\end{tabular}
\caption{Parameter values for our dynamical model of a warped disk precessing around an eccentric binary in the KH 15D system.  Disk inclinations are relative to the binary orbital plane, and disk longitude of ascending nodes are relative to the binary pericenter direction.  See text for definitions and discussion.}
\label{tab:secdyn}
\end{table}

\begin{figure}
    \centering
    \includegraphics[width=\linewidth]{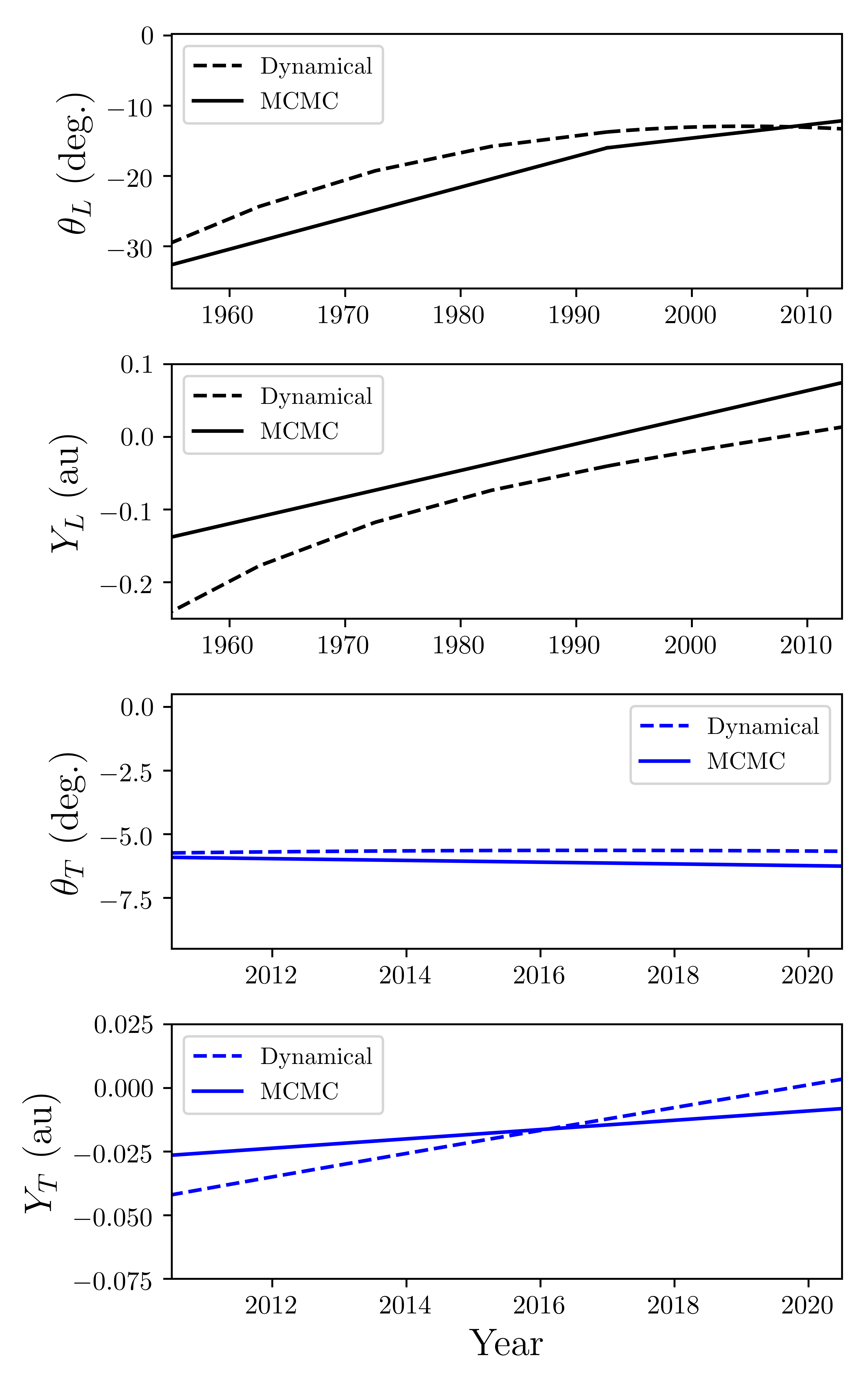}
    \caption{Comparing our dynamical warped disk model (Table~\ref{tab:secdyn}) with the MCMC fits from our phenomenological photometry model (Table~\ref{tab:best_fit_new}).  Although agreement between the two models can be improved, the warped disk reproduces the main features of the photometry model.}
    \label{fig:secdyn}
\end{figure}

Motivated by the estimated leading and trailing estimates in the previous subsection, we experiment with the warped disk orbital parameters and global precession frequency, to find a disk whose properties match the MCMC fitted parameters.  Table~\ref{tab:secdyn} presents example model parameters for a dynamically evolving, warped disk whose features are compatible with the light curve fits, with Figure~\ref{fig:secdyn} displaying $\theta_k(t)$ and $Y_k(t)$ for both (dynamical and MCMC) models over the duration of time the leading and trailing occultations have been observed.  The dynamical model and MCMC fits match one another within a factor of a few, heavily reinforcing the idea that the light curve of KH 15D is caused by a warped, relatively narrow, precessing disk, occulting the starlight of the eccentric binary.  In particular, we see the behavior of the dynamical model matches the $\theta_L(t)$ light curve fit, reproducing the decrease in $\dot \theta_L$ before and after the year $t = t_3 \simeq 1993$.

\begin{figure}
    \centering
    \includegraphics[width=\linewidth]{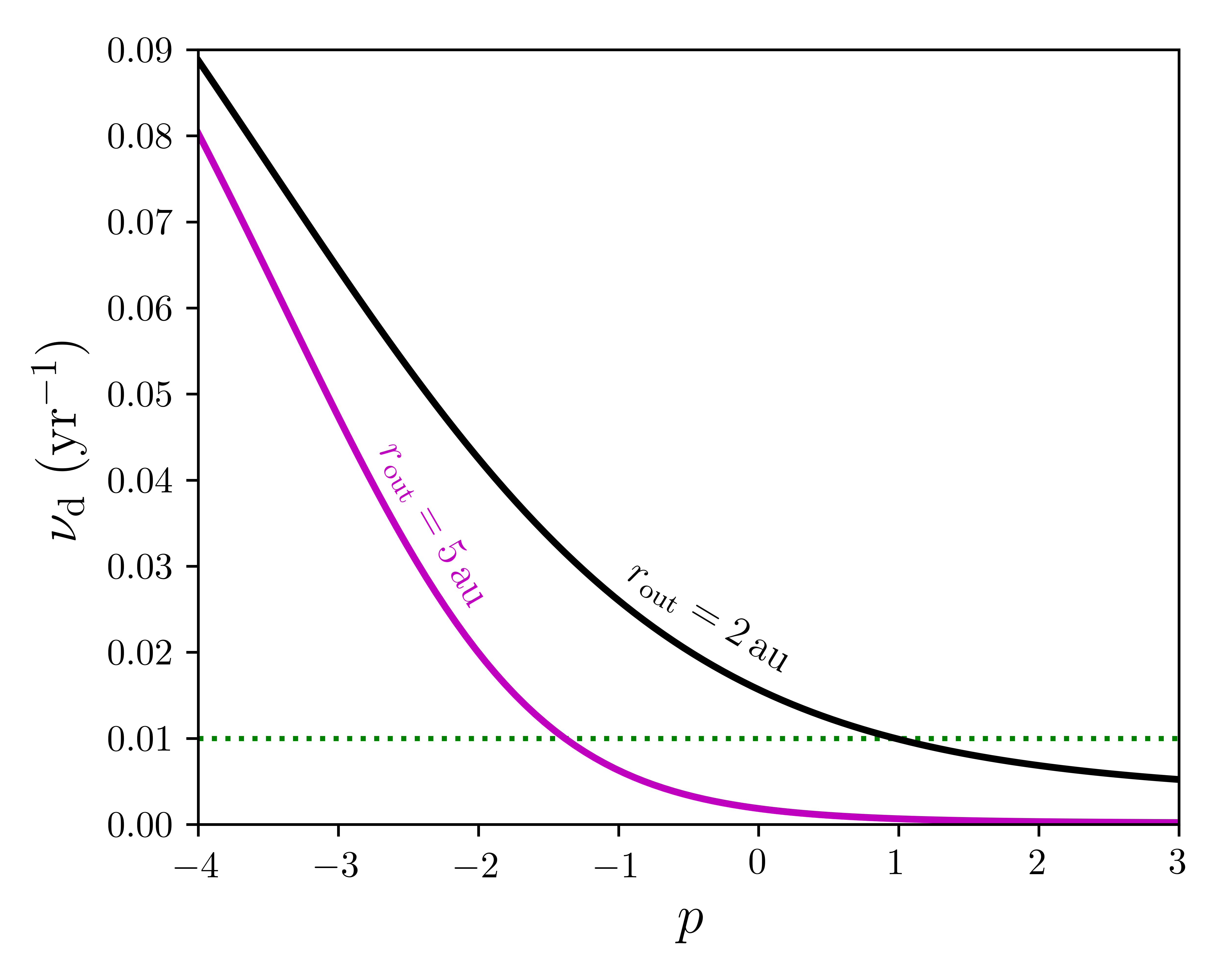}
    \caption{Global disk precession frequnecy $\nu_{\rm d}$ (eq.~\ref{eq:nud_rigid}) as a function of the surface density powerlaw index $p$ ($\Sigma \propto r^p$), for the KH 15D system parameters, assuming $r_{\rm in} = 0.5 \, {\rm au}$ with the $r_{\rm out}$ values indicated.  Dotted green line shows our dynamical model value of $\nu_{\rm d} = 0.01 \, {\rm yr}^{-1}$.  Depending on the disk radial extent, a measurement of $\nu_{\rm d}$ translates to a constraint on $p$.}
    \label{fig:p}
\end{figure}

Complimentary constraints on the disk of KH 15D have come recently from the double-peaked line profile of neutral oxygen emission, assuming the $[{\rm OI}] \ \lambda 6300$ emission originates from the surface of the gaseous circumbinary disk of KH 15D \citep{Fang(2019)}.  This line profile was used to constrain the disk radial extent, as well as the disk surface density profile.  \cite{Fang(2019)} found an inner disk radius of $r_{\rm in} \approx 0.57 \, {\rm au}$, an outer radius $r_{\rm out} \approx 5.2 \, {\rm au}$, and surface density profile $\Sigma \propto r^{-2.9}$.  The inner and outer radii are roughly consistent with our $r_L = r_{\rm in}$ and $r_T = r_{\rm out}$ values constrained by our crude dynamical fit to the photometry of KH 15D (Table~\ref{tab:secdyn}).  We note that the outer edge of a protoplanetary disk gas and dust radius may differ, due to radial drift of the dust \citep[e.g.][]{Weidenschilling(1977),TakeuchiLin(2002),BirnstielAndrews(2014),Powell(2017),Rosotti(2019)}. Indeed, molecular line and continuum emission have been shown to extend to different radii around young stellar objects \citep[e.g.][]{Panic(2009),Andrews(2012),deGregorio-Monsalvo(2013),Ansdell(2018),Facchini(2019)}, showing gas and dust in protoplanetary disks often extend out to different radii (sometimes differing by as much as a factor of $\sim 3$).  The dynamics of dust in a precessing circumbinary disk can also be non-trivial \citep{Poblete(2019),AlyLodato(2020)}. 

Our warped disk model can also constrain the disk surface density profile $\Sigma \propto r^p$, because the distribution of mass within the disk affects the torque exerted on the disk by the binary, modifying the disk precession frequency $\nu_{\rm d}$.  Assuming a nearly-flat disk which is driven into rigid-body precession about the binary, $\nu_{\rm d}$ can be shown to be \citep[e.g.][]{Lodato2013,Foucart2014,Zanazzi2018,LubowMartin(2018)}
\be
\nu_{\rm d} = \frac{3}{4} \left( \frac{5/2 + p}{1 - p} \right) \left[ \frac{1 - (r_{\rm out}/r_{\rm in})^{p-1}}{(r_{\rm out}/r_{\rm in})^{5/2+p} - 1} \right] \frac{\mu}{M_{\rm t}} \left( \frac{a}{r_{\rm in}} \right)^2 \sqrt{ \frac{G M_{\rm t}}{r_{\rm in}^3} }.
\label{eq:nud_rigid}
\ee
Figure~\ref{fig:p} plots the $\nu_{\rm d}$ value given by equation~\eqref{eq:nud_rigid} as a function of $p$.  Depending on the disk outer radius, we clearly see a measurement of $\nu_{\rm d}$ can constrain the $p$ value of the disk.  The model parameters from Table~\ref{tab:secdyn} support $p \sim 1$, which differs substantially from the \cite{Fang(2019)} constraint of $p \approx -2.9$. However, we note that this discrepancy relies on the disk [OI] emission arising from a gaseous disk associated with the occulting ring, as opposed to the interpretation given by \cite{Mundt+(2010)}, who argued the [OI] emission originated from a bipolar jet associated with one or both of the stars at the center of KH 15D. Further photometric modelling is required to see if the $p$ value implied by the disk precession frequency differs from that constrained by the disk ${\rm OI}$ emission.

\section{Theoretical Implications of Dynamical Warped Disk Model}
\label{sec:Discuss}

In \S\ref{sec:Secular}, we showed how the photometry of KH 15D could be explained by a precessing circumbinary disk, in agreement with the results of other works \citep{Winn(2004),Winn2006,Chiang2004,SilviaAgol(2008)}.  The parameters constrained by the photometry of the system are listed in Table~\ref{tab:secdyn}.  Although the fit of the dynamical model to the photometry is crude, we argue the basic conclusions on the parameters of the system are unlikely to differ by more than a factor of a few, and comprises some of the first constraints on small warps within protoplanetary disks.  This section connects the constraints of our dynamical model to theories describing warp propagation in accretion disks, as well as speculation on the long-term evolution of the system.  We also discuss predictions from our model, as well as future modelling efforts.

\subsection{Explaining the Warp and Twist within KH 15D}
\label{sec:Discuss_warp}

Our dynamical model requires a non-zero warp ($\Dg I = I_T - I_L$) and twist ($\Dg \Omega = \Omega_T - \Omega_L$) to cause the complex series of occultations seen in the KH 15D system.  These warps and twists arise from the disk resisting the differential nodal precession induced by the specific torque of the binary
\be
|{\bm T}_{\rm bin}| \sim r^2 n \nu |\bar I|,
\label{eq:T_bin}
\ee
where $n = \sqrt{G M_{\rm t}/r^3}$ is the rings orbital frequency, $\nu$ is the characteristic nodal precession frequency induced on the disk from the binary (eq.~\ref{eq:nu}), and $\bar I$ is the characteristic ``average'' inclination of the disk.  One way to balance the torque from the binary is by thermal pressure between ringlets, which has an internal torque of order \citep[e.g.][]{Ogilvie(1999),ChiangCutler(2003),Chiang2004}
\be
|{\bm T}_{\rm press}| \sim c_{\rm s}^2 |\Dg I|,
\label{eq:T_press}
\ee
where $c_{\rm s} = h r n$ is the ring sound-speed, while $h$ is the aspect ratio of the disk.  Assuming torque balance ($|{\bm T}_{\rm press}| \approx |{\bm T}_{\rm bin}|$) allows us to estimate the warp which may develop under the resisting influence of thermal pressure:
\begin{align}
&\left| \frac{\Dg I}{\bar I} \right|_{\rm press} \sim \left. \frac{r^2 n \nu}{c_{\rm s}^2} \right|_{r=\bar r} 
\nonumber \\
&= 13 \left( \frac{0.05}{h} \right)^2 \left( \frac{\mu}{0.33 \, \Msun} \right) \left( \frac{1.32 \, \Msun}{M_{\rm t}} \right) \left( \frac{a}{0.29 \, {\rm au}} \right)^2 \left( \frac{0.7 \, {\rm au}}{\bar r} \right)^2,
\label{eq:warp_press}
\end{align}
where $\bar r$ is some characteristic radius within the disk.  Clearly, this warp is quite large.  

However, in nearly-inviscid (Shakura-Sunyaev  parameter $\alpha \lesssim h$) disks with the radial-epicyclic frequency satisfying $\kappa^2 \approx n^2$, the near-resonant propagation of bending waves across the disk can amplify the strength of the hydrodynamical torque by a factor \citep{PapaloizouLin(1995),LubowOgilvie(2000),Ogilvie(2006)}
\be
|{\bm T}_{\rm bw}| \sim \frac{1}{|\tilde \kappa|} |{\bm T}_{\rm press}|,
\label{eq:T_bw}
\ee
where $\tilde \kappa \equiv (\kappa^2 - n^2)/(2 n^2)$ is a dimensionless quantity related to the apsidal precession rate.  Because the secular apsidal precession rate is $|\tilde \kappa| \sim \nu/n$ for circumbinary disks \citep{Miranda2015}, torque balance ($|{\bm T}_{\rm bw}| \approx |{\bm T}_{\rm bin}|$) gives
\begin{align}
    &\left| \frac{\Dg I}{\bar I} \right|_{\rm bw} \sim \left. \frac{r^2 \nu^2}{c_{\rm s}^2} \right|_{r=\bar r}
    \nonumber \\
    &= 0.42 \left( \frac{0.05}{h} \right)^2 \left( \frac{\mu}{0.33 \, \Msun} \right)^2 \left( \frac{1.32 \, \Msun}{M_{\rm t}} \right)^2 \left( \frac{a}{0.29 \, {\rm au}} \right)^4 \left( \frac{0.7 \, {\rm au}}{\bar r} \right)^4.
    \label{eq:warp_bw}
\end{align}
This estimate is much closer to the $|\Dg I/\bar I| \sim 1$ values implied by our dynamical model, and lies in agreement with more detailed calculations of warp propagation in protoplanetary disks \citep{Lodato2013,FoucartLai(2013),Foucart2014,Zanazzi2018,LubowMartin(2018)}.

Disk self-gravity can also resist differential nodal precession from the binary.  Mutually-misaligned ringlets experience specific mutual internal torques of order \citep{ChiangCutler(2003),Chiang2004,TremaineDavis(2014),Zanazzi2017,Batygin(2018)}
\be
|{\bm T}_{\rm sg}| \sim \frac{G \Sigma r}{h} |\Dg I| \sim \frac{G M_{\rm d}}{h r} |\Dg I|,
\ee
assuming the disk mass $M_{\rm d} \sim r^2 \Sigma$, and the additional factor of $h^{-1}$ arises from the enhancement of the mutual gravitational attraction between ringlets when the disk is vertically thin \citep{Batygin(2018)}.  Torque balance ($|{\bm T}_{\rm sg}| \approx |{\bm T}_{\rm bin}|$) leads to warps of order
\begin{align}
&\left| \frac{\Dg I}{\bar I} \right|_{\rm sg} \sim \left. \frac{r^3 n \nu h}{G M_{\rm d}} \right|_{r=\bar r}
\nonumber \\
&= 1.3 \left( \frac{h}{0.05} \right)\left( \frac{\mu}{0.33 \, \Msun} \right) \left( \frac{1.7 \, {\rm M}_{\rm Jup}}{M_{\rm d}} \right) \left( \frac{a}{0.29 \, {\rm au}} \right)^2 \left( \frac{0.7 \, {\rm au}}{\bar r} \right)^2 .
\end{align}
Even after assuming the upper limit on the total (gas and dust) disk mass inferred by ALMA observations \citep{Aronow2018}, self-gravity is typically not as effective as bending waves at enforcing coplanarity between ringlets.  However, a massive disk ($M_{\rm d} \sim 1 \, {\rm M}_{\rm Jup}$) can give warps comperable to those inferred by our dynamical KH 15D disk model.

The direction of the warp ($\Dg I$ positive or negative) has also been argued to encode information on the internal forces/torques enforcing disk coplanarity.  \cite{Chiang2004} argued thermal pressure predicts $\Dg I < 0$, while self-gravity predicts $\Dg I > 0$.  More detailed calculations support the prediction that a disk should relax to a $\Dg I > 0$ profile under the influence of disk self-gravity \citep{Batygin(2012),Batygin(2018),Zanazzi2017}.  But calculations taking into account the resonant propagation of bending waves \textit{also} predict $\Dg I > 0$ \citep[e.g.][]{Facchini(2013),Foucart2014,Zanazzi2018,LubowMartin(2018)}.  Hydrodynamical simulations of protoplanetary disks (neglecting self-gravity) find conflicting results, with $\Dg I > 0$ and $\Dg I < 0$ at different times, primarily because the simulations usually cannot be run long enough for the system to relax to a smoothly-evolving warp profile \citep[e.g.][]{Facchini(2013),MartinLubow(2017),MartinLubow(2018),Smallwood(2019),Smallwood(2020),Moody(2019)}.  We note that the disk may never relax to a steady-state.  Simulations which accurately calculate how the binary interacts with a tidally-truncated circumbinary disk find highly-dynamical inner disk edges for disks orbiting eccentric binaries \citep[e.g.][]{Miranda(2017),Munoz(2019),Munoz(2020),Franchini(2019)}.  Because resonant Lindblad torques often truncate disks \citep[e.g.][]{ArtymowiczLubow(1994),LubowMartin(2015),Miranda2015}, which may also excite disk tilts \citep{Borderies(1984),Lubow(1992),ZhangLai(2006)}, it is not unreasonable to say a real circumbinary disk may never relax to a steady-state inclination profile.  We conclude that bending-wave propagation is the main internal force enforcing rigid precession of the disk of KH 15D, despite the conflicting predictions for the sign of $\Dg I$.

A small viscosity in a circumbinary disk also leads to a non-zero twist, due to the azimuthal shear induced by differential nodal precession.  The magnitude of the torque resisting nodal shear is \citep{PapaloizouPringle(1983),PapaloizouLin(1995),Ogilvie(1999),LubowOgilvie(2000)}
\be
|{\bm T}_{\rm visc}| \sim \frac{1}{\alpha} c_{\rm s}^2 |\Dg \Om|,
\label{eq:T_visc}
\ee
assuming an isotropic kinematic viscosity ($\nu = \alpha p/[\rho n]$).  The $\alpha^{-1}$ (rather than $\alpha^{+1}$) dependence in equation~\eqref{eq:T_visc} is from near-resonant forcing of radial and azimuthal perturbations (since $\kappa^2 \approx n^2$), which are damped only by viscosity \citep{PapaloizouLin(1995),LubowOgilvie(2000),LodatoPringle(2007)}.  Viscosity leads to twists of order (assuming $|{\bm T}_{\rm visc}| \approx |{\bm T}_{\rm bin}|$)
\begin{align}
\left| \frac{\Dg \Om}{\bar I} \right|_{\rm visc} &\sim \left. \frac{\alpha r^2 n \nu}{c_{\rm s}^2} \right|_{r=\bar r}
\nonumber \\
= 0.13 & \left( \frac{\alpha}{0.01} \right) \left( \frac{0.05}{h} \right)^2 \left( \frac{\mu}{0.33 \, \Msun} \right) 
\nonumber \\
&\times\left( \frac{1.32 \, \Msun}{M_{\rm t}} \right) \left( \frac{a}{0.29 \, {\rm au}} \right)^2 \left( \frac{0.7 \, {\rm au}}{\bar r} \right)^2.
\label{eq:dOm_visc}
\end{align}
More detailed calculations typically give positive $\Dg \Om$ values a bit larger in circumbinary disks \citep{Foucart2014,Zanazzi2018}, in agreement with our dynamical model.  Although observations frequently infer $\alpha$ values much lower than $10^{-2}$ \citep[e.g.][]{Hughes(2011),Flaherty(2015),Teague(2016),Rafikov(2017),Ansdell(2018)}, the large warp in this disk can excite parametric instabilities, enhancing the viscous dissipation rate in the disk \citep{Goodman(1993),RyuGoodman(1994),Gammie(2000),OgilvieLatter(2013),PaardekooperOgilvie(2019)}.  

The arguments above slightly favor a disk held together by resonant bending waves over self-gravity.  However, such an interpretation requires the scaleheight of the gas be much higher than that of the solids (dust, pebbles, or planetesimals), which must be sufficiently small to cause the sharp occultations seen in the KH 15D light-curve ($H_{\rm solid} \lesssim R_{\rm A}, R_{\rm B}$).  Although no firm detection of disk gas within the KH 15D system has been made, \cite{Lawler+(2010)} detected Na I D line emission and absorption from KH 15D, with a column density which did not vary as the stars became more inclined to the disk midplane.  If the Na I D emission/absorbtion is from the disk gas (not the interstellar medium), this implies a large gas scaleheight ($H_{\rm gas} \gg R_{\rm A}, R_{\rm B}$).  A discrepancy between the gas and solid scaleheights is expected theoretically, as aerodynamical drag causes particles to settle to the disk midplane \citep[e.g.][]{YoudinLithwick(2007)}.  Without gas, dust/solids/planetesimals tend to have larger scaleheights due to mutual gravitational interactions which excite particle inclinations \citep[e.g.][]{Goldriech+(2004)}.  Moreover, if the disk has no gas, because the solids must be optically-thick to starlight, the required solid densities would cause frequent collisions between particles, and imply the KH 15D disk has a short lifetime \citep[e.g.][]{Wyatt+(2008)}.

We conclude the disk warp and twist implied by our model lie in accord with hydrodynamical theories of warped accretion disks.

\subsection{Long-Term Dynamical Evolution of KH 15D}

\begin{figure}
    \centering
    \includegraphics[width=\linewidth]{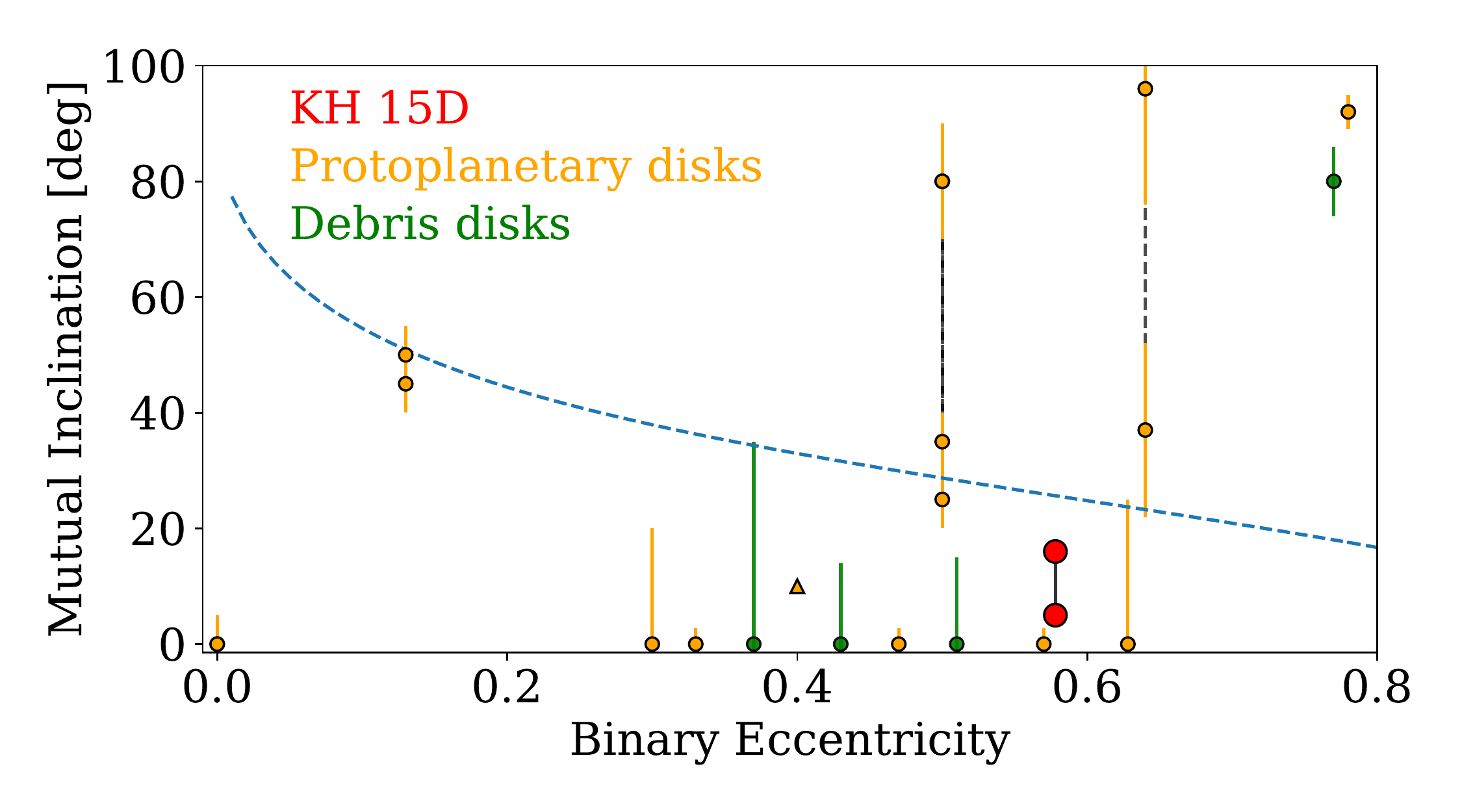}
    \caption{The mutual inclination between the disk and binary orbital plane in the KH 15D system (red), plotted alongside circumbinary disk inclinations for protoplanetary (orange) and debris (green) disks \protect\citep{Czekala2019}, as a function of binary eccentricity. The dashed blue lines plots the critical inclination (eq.~\ref{eq:I_crit}). The black dotted lines connect degenerate solutions for HD 142527, SR 24N and GG Tau Aa-Ab. The triangle represents the lower limit for R CrA.  The disk in KH 15D will align (not polar align) with the orbital plane of the binary.}
    \label{fig:mi_vs_ecc}
\end{figure}

Recently, \cite{Czekala2019} showed circumbinary disks (both gas and debris) have higher inclinations when orbiting eccentric binaries.  Figure~\ref{fig:mi_vs_ecc} displays the disk inclinations analyzed in \cite{Czekala2019}, alongside our constraints for the inclination of KH 15D, which we take directly from our photometric fits ($|\theta_T[t_3]| \lesssim |I_{\rm KH \ 15D}| \lesssim |\theta_L[t_3]|$, see \S\ref{sec:Secular_est}).  The dashed blue line displays the critical inclination \citep{Aly(2015),MartinLubow(2017),Zanazzi2018}
\be
I_{\rm crit} = \cos^{-1} \sqrt{ \frac{5 e^2}{1 + 4e^2} },
\label{eq:I_crit}
\ee
which is a necessary (but not sufficient) condition for the disk-binary inclination to evolve to $90^\circ$ (polar alignment).  From Figure~\ref{fig:mi_vs_ecc}, because $|I_{\rm crit}| > |I_{\rm KH \ 15D}|$, we can be confident the disk will not polar align, and will eventually align with the orbital plane of the binary (without any other mechanisms exciting the disk inclination).

We can also estimate the timescale over which the disk inclination evolves.  Because a non-zero twist $\Dg \Om$ exerts a backreaction torque on the disk from the binary, the disk is driven into alignment (or polar alignment) over the timescale \citep{Foucart2014,Zanazzi2018}
\be
\gamma_{\rm evol} \sim \nu_{\rm d} |\Dg \Om|.
\label{eq:cg_evol_full}
\ee
Inserting equation~\eqref{eq:dOm_visc} into equation~\eqref{eq:cg_evol_full} gives the often-quoted ``Bate timescale'' \citep{Bate2000}.  However, because $\nu_{\rm d}$ and $\Dg \Om$ are both determined by our dynamical model, we can actually estimate $\gamma_{\rm evol}$ using \textit{observationally inferred parameters}:
\be
\gamma_{\rm evol} \sim 2.6 \times 10^{-3} \left( \frac{\nu_{\rm d}}{0.01 \, {\rm yr}^{-1}} \right) \left( \frac{\Dg \Om}{15^\circ} \right) {\rm yr}^{-1}.
\label{eq:cg_evol}
\ee
Equation~\eqref{eq:cg_evol} implies the disk should align with the orbital plane of the binary in less than $\sim 10^3$ years.  Although secular interactions can keep the disk misaligned with the eccentric orbital plane of the binary over timescales a few times longer than estimate~\eqref{eq:cg_evol} \citep{Zanazzi2018, Smallwood(2019)}, this is much shorter than the $\sim 10^6$ year lifetimes of typical protoplanetary disks \citep[e.g.][]{Haisch(2001)}.  Either we are observing KH 15D while it is still very young, or additional mechanisms are exciting the disk inclination.

\subsection{KH 15D Model Predictions and Improvements}

The most immediate consequences are predictions for future light curve behavior from our photometry model fits (Figs.~\ref{fig:unfolded} \&~\ref{fig:folded2}, Table~\ref{tab:best_fit_new}).  Current $I$-band measurements should show the light from star A slowly being revealed by the trailing edge (since $t_7 \approx 2021$).  By the year $\sim$2029, the orbit of star B should be completely revealed, resulting in a ceasing of the variability from this star.  By the year $\sim$2041, we should cease to see photometric variability due to the circumbinary disk.  While our current model which produces a reasonable fit to the photometric data employs an opaque screen with a constant $\dot{\theta}_T$, our dynamical model predicts that the fit can be further improved if the change of $\dot{\theta}_T$ with time is incorporated (eq.~\ref{eq:dtheta_app}). We provide in Table~\ref{tab:online_data} the predicted light curve (in $I$ band) of this system until the year 2050.

We are able to make an explicit connection between the phenomenological model (\S\ref{sec:model}), and a precessing, warped disk occulting the binary of KH 15D (\S\ref{sec:Secular}).  More stringent constraints on the disk geometry would use the inner and outer disk orbital parameters $\{r_k, I_k, \Om_k\}$ and global disk precession frequency $\nu_{\rm d}$, rather than parameters describing the locations and orientations of the leading and trailing edges $\{\theta_k, \dot \theta_k, Y_k, \dot Y_k\}$, to fit the light curve of KH 15D.  This exercise should yield parameters consistent with those listed in Table~\ref{tab:secdyn} within a factor of a few.

Our folded light curves (Figs~\ref{fig:folded1}-\ref{fig:folded2}) show the leading edge is well fit by a sharp edge, whereas the poor fit for the trailing edge imply it is clumpy/puffy, in agreement with the findings of \cite{GarciaSoto2020}.  The sharp inner edge is likely due to tidal truncation by the torque from the binary.  Calculations and hydrodynamical simulations suggest that the radius at which the binary truncates the disk is $\sim$2 times the binary semi-major axis ($a \approx 0.3$ au) (e.g. \citealt{Miranda2015}), lying close to the inner radius value of our dynamical model (Table~\ref{tab:secdyn}).  However, it remains unclear why the disk is so compact ($r_{\rm out} \lesssim 5 \ {\rm au}$), and the possibility still exists the disk outer edge is truncated by a planet \citep{Chiang2004}. There exists tentative observational support for this hypothesis, as \cite{Arulanantham+(2017)} found infrared-excess from KH 15D consistent with the thermal emission from a $\sim$10 $M_{\rm Jup}$ mass planet. Future modelling of how dust scattering and the finite optical depth of the disk can create a ``fuzzier'' outer edge would be of interest \citep{Chiang2004,SilviaAgol(2008)}.

\section{Conclusions}
\label{sec:Conc}

In this work, we have developed a circumbinary disk model that explains the photometric variability of KH 15D spanning more than 60 years.  From this model, we are able to constrain the disk annular extent, inclination, orientation with respect to the binary pericenter direction, warp profile, precession frequency, and even surface density profile.  The fits of our phenomenological model to fit the photometry of KH 15D are displayed in Table~\ref{tab:best_fit_new}, with parameters of a warped disk which are consistent with the phenomenological model constraints listed in Table~\ref{tab:secdyn}.  Although strict constraints on the warped disk remain elusive, we can be confident about the following features of the disk:
\begin{itemize}
\item The beginning of the dips/occultations in KH 15D are due to the disk inner edge slowly covering the binary, while the currently observed slow reversal of the dipping behavior in KH 15D is due to the disk outer edge slowly revealing the binary.  The inner edge has a radius $r_{\rm in} \lesssim 1 \ {\rm au}$, while the outer edge has a radius $r_{\rm out} \sim {\rm few} \ {\rm au}$.
\item The disk inner edge is more inclined to the orbital plane of the binary than the disk outer edge.  Both inner and outer disk inclinations are less than $\sim 16^\circ$, but greater than $\sim 5^\circ$, with a difference of order $\sim 10^\circ$.
\item The disk inner and outer longitude of ascending nodes differ by $\sim 15^\circ$.
\end{itemize}
These constraints are consistent with hydrodynamical theories of warped accretion disks, resisting differential nodal precession from the gravitational torque from the binary (\S\ref{sec:Discuss_warp}).

Our models also find a precessional period of order $P_{\rm prec} \sim 2\pi/\nu_{\rm d} \sim 600 \ {\rm years}$, but this constraint is sensitive to the model fit of KH 15D.  We can be very confident, however, that the timescale over which the disk of KH 15D aligns with the orbital plane of the binary is much shorter than the lifetime of the disk (eq.~\ref{eq:cg_evol}), suggesting that additional mechanisms are exciting the disk tilt.

\section*{Acknowledgements}

We thank Eugene Chiang, Ruth Murray-Clay, and Norm Murray for useful discussions, and Jessica Speedie for help in developing the circumbinary disk visualization in \autoref{fig:model1}. We would also like to thank the referee, William Herbst, for comments and suggestions that have improved the quality of this work. MP has been supported by undergraduate research fellowships from the Centre for Planetary Sciences and Canadian Institute for Theoretical Astrophysics (CITA). JZ and WZ were supported by the Natural Sciences and Engineering Research Council of Canada (NSERC) under the funding reference \# CITA 490888-16.

\section*{Data Availability}

The data underlying this article are available in the public domain \citep{Hamilton(2003),Johnson2004,JohnsonWinn(2004),Maffei2005,Winn2006,Aronow2018,GarciaSoto2020}. The data produced in this work are available in its online supplementary material (Table~\ref{tab:online_data}).


\bibliographystyle{mnras}
\bibliography{Poon+_v4} 


\appendix

\section{Model Parameters}

We display all MCMC parameter fits for our new photometric model, for various $\alpha$, as described in Section 2.2. Recall $\alpha$ is the ratio of the trail edge velocity over the lead edge velocity along the vertical axis of our line of sight. Small $\alpha$ tests for narrow disk models whereas large $\alpha$ tests for extended disk models. We do not report model parameters for $\alpha=10.0$ since the MCMC does not converge. Upper and lower error bars indicate a $1\sigma$ confidence interval. Model parameters are described in \autoref{tab:description}.

\begin{table*}
\centering
\caption{Same as Table~\ref{tab:best_fit_new}, except we vary the value of $\alpha$.}
\label{tab:best_fit_new2}
\begin{tabular}{lcccc}
\hline
    Free parameter & $\alpha=0.1$ & $\alpha=0.3$ & $\alpha=2.0$  & $\alpha=3.0$ \\ \hline
    
    $P$ [days] & $48.3786 \substack{+0.0002 \\ -0.0002}$ & $48.3781 \substack{+0.0002 \\ -0.0002}$ & $48.3713 \substack{+0.0002 \\ -0.0002}$ & $48.3691 \substack{+0.0002 \\ -0.0001}$\\
    $e$ & $0.5794 \substack{+0.0008 \\ -0.0008}$ & $0.5771 \substack{+0.0009 \\ -0.0009}$ & $0.5716 \substack{+0.0008 \\ -0.0008}$ & $0.5677 \substack{+0.0008 \\ -0.0008}$\\
    $I$ [deg] & $91.002 \substack{+0.003 \\ -0.001}$ & $91.002 \substack{+0.003 \\ -0.001}$ & $91.0004 \substack{+0.0008 \\ -0.0003}$ & $91.0005 \substack{+0.0008 \\ -0.0003}$\\
    $\omega$ [deg] & $9.80 \substack{+0.05 \\ -0.06}$ & $11.29 \substack{+0.06 \\ -0.06}$ & $12.04 \substack{+0.06 \\ -0.06}$ & $11.97 \substack{+0.06 \\ -0.07}$\\
    $T_p$ [JD] - 2,452,350 & $3.48 \substack{+0.02 \\ -0.01}$ & $4.00 \substack{+0.02 \\ -0.02}$ & $4.32 \substack{+0.02 \\ -0.02}$ & $4.31 \substack{+0.02 \\ -0.02}$\\
    $L_B/L_A$ & $1.63 \substack{+0.01 \\ -0.02}$ & $1.57 \substack{+0.02 \\ -0.01}$ & $1.17 \substack{+0.01 \\ -0.01}$ & $1.064 \substack{+0.009 \\ -0.009}$\\
    $\epsilon_1$ & $0.0516 \substack{+0.0008 \\ -0.0008}$ & $0.0481 \substack{+0.0007 \\ -0.0007}$ & $0.071 \substack{+0.001 \\ -0.001}$ & $0.085 \substack{+0.001 \\ -0.001}$\\
    $\epsilon_2$ & $0.097 \substack{+0.001 \\ -0.001}$ & $0.0651 \substack{+0.0009 \\ -0.0009}$ & $0.0495 \substack{+0.0006 \\ -0.0007}$ & $0.0499 \substack{+0.0007 \\ -0.0007}$\\
    $\xi_1$ & $1.81 \substack{+0.03 \\ -0.04}$ & $1.44 \substack{+0.02 \\ -0.02}$ & $2.40 \substack{+0.04 \\ -0.04}$ & $3.18 \substack{+0.07 \\ -0.06}$\\
    $\xi_2$ & $7.5 \substack{+0.2 \\ -0.2}$ & $3.60 \substack{+0.07 \\ -0.06}$ & $2.97 \substack{+0.05 \\ -0.05}$ & $3.04 \substack{+0.05 \\ -0.05}$\\
    $t_3$ & $1992.70 \substack{+0.06 \\ -0.06}$ & $1993.16 \substack{+0.05 \\ -0.05}$ & $1991.49 \substack{+0.07 \\ -0.08}$ & $1991.3 \substack{+0.1 \\ -0.1}$\\
    $t_5$ & $2008.16 \substack{+0.01 \\ -0.01}$ & $2008.12 \substack{+0.01 \\ -0.01}$ & $2007.72 \substack{+0.01 \\ -0.01}$ & $2007.76 \substack{+0.02 \\ -0.02}$\\
    $t_6$ & $2013.18 \substack{+0.03 \\ -0.03}$ & $2013.46 \substack{+0.03 \\ -0.03}$ & $2012.44 \substack{+0.02 \\ -0.02}$ & $2012.07 \substack{+0.02 \\ -0.02}$\\
    $\theta_L(t_3)$ [deg] & $-55 \substack{+1 \\ -1}$ & $-20.0 \substack{+0.3 \\ -0.3}$ & $-13.5 \substack{+0.2 \\ -0.2}$ & $-13.7 \substack{+0.2 \\ -0.2}$\\
    $\theta_T(t_3)$ [deg] & $-2.7 \substack{+0.4 \\ -0.4}$ & $-2.4 \substack{+0.2 \\ -0.2}$ & $-21.0 \substack{+0.3 \\ -0.3}$ & $-29.7 \substack{+0.4 \\ -0.5}$\\
    $\dot{\theta}_{L1}$ [rad/year] & $0.0075 \substack{+0.0004 \\ -0.0005}$ & $0.0099 \substack{+0.0002 \\ -0.0003}$ & $0.0025 \substack{+0.0002 \\ -0.0002}$ & $0.0019 \substack{+0.0003 \\ -0.0003}$\\
    $\dot{\theta}_{L2}$ [rad/year] & $0.0093 \substack{+0.0003 \\ -0.0003}$ & $0.0033 \substack{+0.0002 \\ -0.0002}$ & $0.0042 \substack{+0.0001 \\ -0.0001}$ & $0.0046 \substack{+0.0001 \\ -0.0001}$\\
    $\dot{\theta}_T$ [rad/year] & $-0.0035 \substack{+0.0003 \\ -0.0003}$ & $-0.0030 \substack{+0.0001 \\ -0.0001}$ & $0.0065 \substack{+0.0002 \\ -0.0002}$ & $0.0088 \substack{+0.0003 \\ -0.0003}$\\ \hline
    Fit photometry? & No & Yes & No & No \\
    $\chi^2_{\rm phot}$ & 16386 & 13822 & 15558 & 17222 \\
    $\chi^2_{\rm RV}$ & 12 & 13 & 16 & 17 \\
    $\text{Reduced } \chi^2$ & 1.62 & 1.40 & 1.60 & 1.76 \\
\hline
\end{tabular}
\end{table*}

\section{MCMC Corner Plots}

We display the corner plots to our best fit model ($\alpha=0.5$) with best fit values listed in \autoref{tab:best_fit_new}. We remove the first 17,500 of 20,000 total steps as burn-in, and plot the posterior distribution. The apparent degeneracy with $\xi_2$ and $L_B$ appears in many of the MCMC fits. This is likely due to some subtleties of the halo model, yet they do not affect the quality of the photometric fits.  Because we are primarily interested in constraints on the properties describing the ascent of the leading and trailing screens $\{\theta_k, \dot \theta_k, Y_k, \dot Y_k\}$, to be consistent with \cite{Winn2006}, we do not modify the halo model. 

\begin{figure*}
    \centering
    \includegraphics[width=\linewidth]{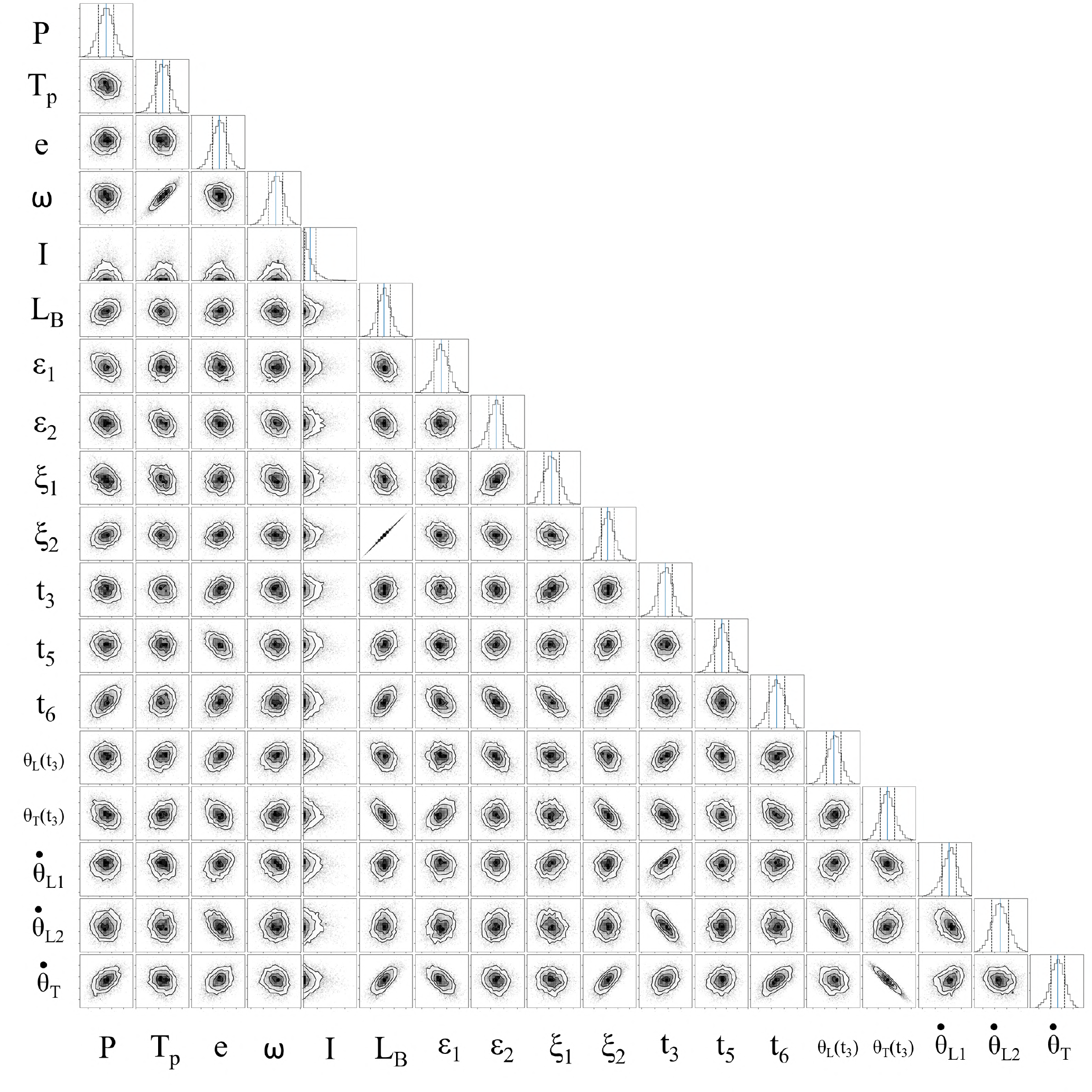}
    \caption{Two dimensional projection of the posterior probability distribution sampled using MCMC for $\alpha=0.5$. Blue solid lines indicate best fit values reported in \autoref{tab:best_fit_new}, whereas black dashed lines indicate a $1\sigma$ confidence interval.}
    \label{fig:corner}
\end{figure*}

\bsp	
\label{lastpage}
\end{document}